# Enhancing the SST Turbulence Model with Symbolic Regression: A Generalizable and Interpretable Data-Driven Approach

Chenyu Wu[1], Yufei Zhang[1,*]

(1. School of Aerospace Engineering, Tsinghua University, Beijing, 100084, China)

**Abstract:** Turbulence modeling within the Reynolds averaged Navier-Stokes (RANS) equations' framework is essential in engineering due to its high efficiency. Field inversion and machine learning (FIML) techniques have attempted to improve RANS turbulence models' predictive capabilities for separated flows. However, FIML-generated models often lack interpretability, limiting physical understanding and manual improvements based on prior knowledge. Additionally, these models typically struggle with generalization in flow fields distinct from the training set. This study addresses these issues by employing symbolic regression (SR) to derive an analytical relationship between the correction factor of the baseline turbulence model and local flow variables, enhancing the baseline model's ability to predict separated flow across diverse test cases. The shear-stress-transport (SST) model undergoes field inversion on a curved backward-facing step (CBFS) case to obtain the corrective factor field $\beta$, and SR is used to derive a symbolic map between local flow features and $\beta$. The SR-derived analytical function is integrated into the original SST model, resulting in the SST-SR model. The SST-SR model's generalization capabilities are demonstrated by its successful predictions of separated flow on various test cases, including 2D-bump cases with varying heights, periodic hill case where separation is dominated by geometric features, and the three-dimensional Ahmed-body case. In these tests, the model accurately predicts flow fields, showing its effectiveness in cases completely different from the training set. The Ahmed-body case, in particular, highlights the model's ability to predict the three-dimensional massively separated flows. When applied to a turbulent boundary layer with $Re_L = 1.0 \times 10^7$, the SST-SR model predicts wall friction coefficient and log layer comparably to the original SST model, maintaining the attached boundary layer prediction performance.

* Corresponding author, E-mail: zhangyufei@tsinghua.edu.cn

# 1. Introduction

Accurate simulation of flow turbulence in CFD plays a crucial role in many engineering applications. In airfoil design, understanding turbulence helps engineers minimize flow separation and reduce drag. In the field of automobile engineering, accurate turbulence prediction aids in the development of streamlined body shapes, controlling the vortices formed at the rear of the car. Direct numerical simulation (DNS), large-eddy simulation (LES) and Reynolds averaged Navier-Stokes equations (RANS) are typical ways to simulate the flow turbulence. DNS and LES can give a high-fidelity prediction of complex turbulent flows, but the computational cost is quite high. On the contrary, RANS simulation is fairly cheap and efficient, making it preferable for engineering design applications where dozens of different configurations should be evaluated by CFD in a limited amount of time.

However, the RANS models frequently fail in complex turbulent flows, especially in the prediction of flow separation. In periodic hills, widely used RANS models such as the shear-stress-transport (SST) model and Spalart-Allmaras (SA) model predict an erroneously delayed reattachment point [1]. Similar problems can be seen in the curved backward-facing step (CBFS)[2], NASA hump [3], and the iced airfoil [4]. In three-dimensional cases such as the Ahmed body, typical RANS models also fail in predicting the complex three-dimensional vortices formed by separated flows in the wake of a blunt body [5], which is of particular interest in automobile aerodynamics.

Efforts have been made to improve RANS models' ability to predict separated flows, expanding their applications in engineering design. Some researchers have derived correction terms for existing turbulence models based on fundamental turbulence laws and astute physical insights. Rumsey, for instance, argued that delayed reattachment of separated flow results from underpredicted turbulence activity in the separated shear layer where non-equilibrium turbulence dominates [1]. By adding an analytical correction factor to the destruction term in $\omega$'s equation, Rumsey's SST-sf model achieved superior predictions in various 2D separated flows. Similarly, Li et al. [4][6][7] derived a novel correction term for the $k - \overline{v^2} - \omega$ turbulence model, based on Rumsey's observation, resulting in a model that accurately predicts separated flows on a wide range of iced airfoils and wings. Although these models have clear physical meaning and broad application, deriving correction terms solely from physical arguments can be challenging and heavily reliant on the modeler's experience, potentially introducing bias and hindering the discovery of new features that could enhance model performance.

Data-driven approaches, including uncertainty quantification and machine learning, have recently been applied to turbulence modeling. In uncertainty quantification, Xiao et al. [8][9] used the Ensemble Kalman filter to estimate $k-\omega$ model's Reynolds stress error based on high-fidelity DNS velocity data. Duraisamy et al. [10]-[13] employed an optimization-based field-inversion technique to obtain the multiplicative discrepancy term $\beta(x)$ in the transport equations' destruction and production terms. The optimized $\beta(x)$ distribution minimizes errors between predicted quantities of interest (QoI) and high-fidelity data-derived QoI. These methods allow efficient, formal error quantification in existing models without heavy reliance on physical arguments or empirical observations, enabling error correction proposals for improved performance. Machine learning is now commonly used to map flow features to correction terms. Yan et al. [14][15] utilized artificial neural networks (ANN) within the field inversion and machine learning (FIML) framework to map local flow features to the multiplicative discrepancy term $\beta(x)$. Yin et al. [16] proposed a novel set of input features for ANN model to predict Reynolds stress, which achieved good results in periodic hills. Yin et al. [17] also developed an innovative iterative data-driven turbulence modeling framework using the Random Forest (RF) model. Additionally, probabilistic models like Gaussian process regression predict the multiplicative discrepancy term with uncertainty [18].

Although these models excel in test cases similar to their training set, they struggle with generalization. For instance, FIML-generated models often exhibit unwanted behavior in entirely different test cases, as argued by Rumsey et al. [19], even failing in predicting simple turbulent boundary layer flows. Additionally, machine-learning models, with their thousands of parameters, lack physical interpretability, resembling a black box to users. This makes it difficult to incorporate physical a priori knowledge during training and impedes post-training improvements based on physical insight. Furthermore, these models lack portability, hindering their use in different solvers from where they were trained. Recognizing these limitations, Spalart [20] called for a refined framework to develop more universal and portable RANS turbulence models from data.

Recently, symbolic regression (SR), a classic machine-learning method, has been introduced to physical applications, offering a distinct advantage over traditional turbulence modeling. Unlike

ANN, RF, and GEP models, SR distills a list of compact analytical expressions with varying complexity between inputs and outputs from datasets, rather than relying on black box models with excessive parameters. Some SR frameworks [21]-[23] allow modelers to assign input features and element functions, reflecting their physical understanding of output properties such as boundedness and monotonicity. Final expressions can be discovered using classic genetic algorithms or novel deep-learning-based [24] frameworks.

Applying SR to turbulence modeling enables the generation of short symbolic expressions for correction terms with different complexities. Strong inductive bias and a priori physical knowledge can be injected through input features and element functions, potentially yielding expressions with better generalization ability than black box models like ANN. Compared to traditional turbulence modeling, which relies on pure physical argument, SR leverages data-driven techniques to quantify and source errors in existing models while still resulting in short correction term expressions. SR has been used for discovering algebraic Reynolds-Stress models [25] and nonlinear eddy viscosity relations in multi-phase flows [26]. However, in these works, the SR-derived expressions have a relatively simple polynomial form, and their potential generalizability remains underexplored.

In this study, we present a novel data-assisted turbulence modeling framework to generate an interpretable and generalizable analytical correction term to enhance the baseline model's ability to predict turbulent separated flows. We call this framework FISR (Field Inversion and Symbolic Regression). First, the field inversion is performed on a curved backward-facing step (CBFS) to derive the multiplicative correction term $\beta$ of the SST model. Then, SR is performed on the CBFS dataset to generate an analytical expression of the correction term $\beta$. The expression is then interpreted and partly modified based on our physical knowledge of non-equilibrium turbulence. The expression $\beta(\mathbf{w})$ is then integrated into the code of the SST model to get the SST-SR model. The SST-SR model is then applied to test cases completely different from the training set (CBFS), including 2D bumps with various heights, periodic hills, and 3D Ahmed body. The results show that the SST-SR model outperforms the baseline SST model in all cases, indicating its generalizability. The model also gives similar results as the baseline model for a simple turbulent boundary layer with $Re_L = 1.0 \times 10^7$, showing that the correction term does not affect the baseline model's performance in simple flows.

# 2. The framework of field inversion and symbolic regression

In this study, field inversion is used to derive the $\beta$ field dataset and SR is used to generate the expression of $\beta$ with respect to flow features, as is shown in Figure 1. The outline of the field inversion and the symbolic regression (SR) techniques will be introduced in the following section.

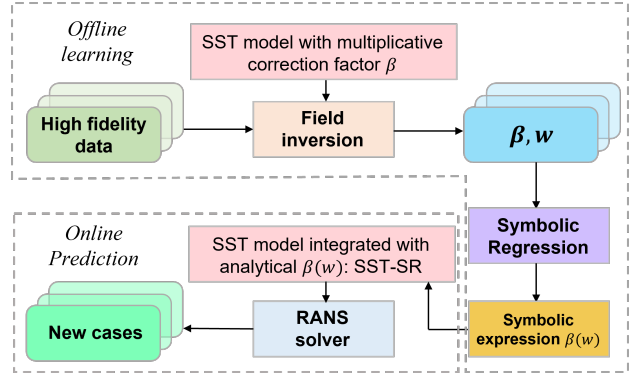

Figure 1. The framework of field inversion and symbolic regression

## 2.1. Field inversion

The shear stress transport (SST) model [27] is widely used in industrial applications. It contains one transport equation for turbulent kinetic energy $k$ and one for specific dissipation rate $\omega$:

$$\frac{\partial(\rho k)}{\partial t} + \frac{\partial(\rho u_j k)}{\partial x_j} = P - \beta^* \rho \omega k + \frac{\partial}{\partial x_j}\left[(\mu + \sigma_k \mu_t)\frac{\partial k}{\partial x_j}\right]$$
$$\frac{\partial(\rho \omega)}{\partial t} + \frac{\partial(\rho u_j \omega)}{\partial x_j} = \frac{\gamma}{\nu_t} P - \theta \rho \omega^2 + \frac{\partial}{\partial x_j}\left[(\mu + \sigma_\omega \mu_t)\frac{\partial \omega}{\partial x_j}\right] + 2(1-F_1)\frac{\rho \sigma_{\omega 2}}{\omega}\frac{\partial k}{\partial x_j}\frac{\partial \omega}{\partial x_j} \quad (1)$$

The production term $P$ is defined as:

$$P = \min(\tau_{ij}\frac{\partial u_i}{\partial x_j}, 10\beta^* \rho k \omega)$$
$$\tau_{ij} = \mu_t\left(2S_{ij} - \frac{2}{3}\frac{\partial u_k}{\partial x_k}\delta_{ij}\right) - \frac{2}{3}\rho k \delta_{ij} \quad (2)$$
$$S_{ij} = \frac{1}{2}\left(\frac{\partial u_i}{\partial x_j} + \frac{\partial u_j}{\partial x_i}\right)$$

Turbulent kinetic eddy viscosity can be computed from $k$ and $\omega$:

$$\mu_T = \frac{\rho a_1 k}{\max(a_1 \omega, SF_2)} \quad (3)$$

$S = \sqrt{2S_{ij}S_{ij}}$. Other functions involved in Eq (1) are defined as follows:

$$\begin{aligned}
F_1 &= \tanh(\xi_1^4) \\
\xi_1 &= \min\left[\max\left(\frac{\sqrt{k}}{\beta^*\omega d}, \frac{500\nu}{d^2\omega}\right), \frac{4\rho\sigma_{\omega 2}k}{CD_{k\omega}d^2}\right] \\
CD_{k\omega} &= \max\left(2\rho\sigma_{\omega 2}\frac{1}{\omega}\frac{\partial k}{\partial x_j}\frac{\partial \omega}{\partial x_j}, 10^{-10}\right) \\
F_2 &= \tanh(\xi_2^2) \\
\xi_2 &= \max\left(2\frac{\sqrt{k}}{\beta^*\omega d}, \frac{500\nu}{d^2\omega}\right)
\end{aligned} \tag{4}$$

Wall distance $d$ at a point $(x, y, z)$ is defined as the distance between the point and the nearest wall. $\gamma_1, \gamma_2, \sigma_{k_1}, \sigma_{k_2}, \sigma_{\omega_1}, \sigma_{\omega_2}, \theta_1, \theta_2, \beta^*, a_1$ [27] are model constants. All the constants in Equation (1) should be blended by the function $F_1$:

$$\phi = F_1\phi_1 + (1-F_1)\phi_2 \tag{5}$$

The SST model is famous for its robustness and has been applied to a wide range of scenarios. However, the model's ability to predict separated flows is limited. Corrections have been made to enhance the SST model's prediction ability, such as the SST-sf model [1] (for separated flow) and SST-RC-Hellsten [28] (for rotation and curvature effect). In these works, the correction term is usually multiplied by the destruction term of the $\omega$'s transport equation:

$$-\theta\rho\omega^2 \rightarrow -f(\mathbf{w})\theta\rho\omega^2 \tag{6}$$

$f(\mathbf{w})$ is a correction factor and $\mathbf{w}$ represents some flow field features such as strain rate, rotation rate, and so on. Since our goal is also to improve the baseline SST model's ability in predicting the separated flow, following the strategy in [1], a spatially distributed multiplicative correction factor $\beta(\mathbf{x})$ is added to the destruction term of $\omega$'s transport equation:

$$\frac{\partial(\rho\omega)}{\partial t} + \frac{\partial(\rho u_j\omega)}{\partial x_j} = \frac{\gamma}{\nu_t}P - \beta\cdot\theta\rho\omega^2 + \frac{\partial}{\partial x_j}\left[(\mu+\sigma_\omega\mu_t)\frac{\partial\omega}{\partial x_j}\right] + 2(1-F_1)\frac{\rho\sigma_{\omega 2}}{\omega}\frac{\partial k}{\partial x_j}\frac{\partial \omega}{\partial x_j} \tag{7}$$

Then we can adjust $\beta$'s distribution to make the quantity of interest (QoI, i.e., velocity) predicted by the SST model match the high-fidelity data (i.e., experiment or DNS). In this study, the optimum $\beta$ distribution is obtained by solving the optimization problem below (in discrete form):

$$\min_{\beta} J = \lambda_{obs}\sum_i(d_i - h_i(\boldsymbol{\beta}))^2 + \lambda_{prior}\sum_j(\beta_j - 1)^2 \tag{8}$$

$\beta_j$ is the value of $\beta$ on the $j$-th grid cell in CFD computation, $\boldsymbol{\beta}$ is a vector whose $j$-th element is $\beta_j$, $d_i$ is the $i$-th QoI from high-fidelity data, $h_i(\boldsymbol{\beta})$ is the $i$-th QoI predicted by the SST model with $\beta$ distribution corresponding to $\boldsymbol{\beta}$, and $\lambda_{obs}, \lambda_{prior}$ are constants. The first term means that we attempt to minimize the error between the predicted QoI and the high-fidelity data. The second term means that we do not want $\beta$ to diverge too far from its original value in the baseline model, 1. $\lambda_{obs} \approx \left[\sum_i (d_i - h_i(1))^2\right]^{-1}$ so that the first term is approximately 1 when $\beta_j = 1, \forall j$. $\lambda_{prior}$ reflects the trade-off between minimizing the error of QoI and preserving the smoothness of the $\beta$ distribution. The larger $\lambda_{prior}$ is, the smoother the $\beta$ distribution is and the larger the QoI error is. $\lambda_{prior}$ is often set to a value around $1 \times 10^{-6} \sim 1 \times 10^{-3}$.

In this paper, we use the gradient-based optimization program SNOPT [29] to solve the problem defined by Equation (8). The adjoint method, which is suitable for optimization problems involving partial differential equations, is used to compute the gradient of the objective function $J$ [30]. Specifically, the discrete adjoint method is used in this paper due to its flexibility [31]. The algorithm of the discrete adjoint method is introduced in the following paragraph.

In CFD computations, the objective function $J$ in Equation (8) and the residual of the discretized governing equations $\boldsymbol{R}$ (note that $\boldsymbol{R}$ is a vector valued function, and the dimension of it is approximately $m \cdot N$, with $m$ representing the number of governing equations and $N$ standing for the total number of grid cells) can be abstractly written as the functions of $\boldsymbol{\beta}$ and flow variable $\boldsymbol{w}$ (which is also a vector):

$$J(\boldsymbol{w}, \boldsymbol{\beta}), \boldsymbol{R}(\boldsymbol{w}, \boldsymbol{\beta}) \tag{9}$$

But in fact, $\boldsymbol{w}$ is also a function of $\boldsymbol{\beta}$ defined by the implicit relation:

$$\boldsymbol{R}(\boldsymbol{w}(\boldsymbol{\beta}), \boldsymbol{\beta}) = 0 \tag{10}$$

Equation (10) simply states that for every $\boldsymbol{\beta}$, the corresponding flow field variable $\boldsymbol{w}$ is a converged solution of the discretized governing equations $\boldsymbol{R}$. Applying chain rules to Equation (10), we get:

$$\frac{\partial \boldsymbol{R}}{\partial \boldsymbol{w}} \frac{d\boldsymbol{w}}{d\boldsymbol{\beta}} + \frac{\partial \boldsymbol{R}}{\partial \boldsymbol{\beta}} = 0 \tag{11}$$

By multiplying the inverse of the square matrix $\partial \boldsymbol{R}/\partial \boldsymbol{w}$ (note that the dimension of $\boldsymbol{R}$ should be equal to the dimension of $\boldsymbol{w}$, otherwise the discrete governing equations cannot be solved), we can arrive at the differential equation system satisfied by the function $\boldsymbol{w}(\boldsymbol{\beta})$:

$$\frac{d\boldsymbol{w}}{d\boldsymbol{\beta}} = -\left(\frac{\partial \boldsymbol{R}}{\partial \boldsymbol{w}}\right)^{-1} \frac{\partial \boldsymbol{R}}{\partial \boldsymbol{\beta}} \tag{12}$$

So, the gradient of $J$ can be expressed by using chain rules and equation (12):

$$\nabla J = \frac{\partial J}{\partial \boldsymbol{\beta}} + \frac{\partial J}{\partial \boldsymbol{w}} \frac{d\boldsymbol{w}}{d\boldsymbol{\beta}} = \frac{\partial J}{\partial \boldsymbol{\beta}} - \frac{\partial J}{\partial \boldsymbol{w}} \left(\frac{\partial \boldsymbol{R}}{\partial \boldsymbol{w}}\right)^{-1} \frac{\partial \boldsymbol{R}}{\partial \boldsymbol{\beta}} \tag{13}$$

If we define the adjoint variable $\boldsymbol{\psi}$ as follows:

$$\left(\frac{\partial \boldsymbol{R}}{\partial \boldsymbol{w}}\right)^T \boldsymbol{\psi} = -\left(\frac{\partial J}{\partial \boldsymbol{w}}\right)^T \tag{14}$$

Then equation (13) can be written as:

$$\nabla J = \frac{\partial J}{\partial \boldsymbol{\beta}} + \boldsymbol{\psi}^T \frac{\partial R}{\partial \boldsymbol{\beta}} \tag{15}$$

The discrete adjoint method includes the following steps:

1. After the CFD computation is converged (i.e., $\boldsymbol{R}(\boldsymbol{w}(\boldsymbol{\beta}), \boldsymbol{\beta}) = 0$ is satisfied), $\left(\frac{\partial \boldsymbol{R}}{\partial \boldsymbol{w}}\right)^T$ and $\left(\frac{\partial J}{\partial \boldsymbol{w}}\right)^T$ are obtained by using automatic differentiation [32].

2. The linear equation (14) is solved to acquire the adjoint variable $\boldsymbol{\psi}$

3. The gradient of the objective function $J$ is computed by using equation (15).

In this paper, the RANS solver and the discrete adjoint solver are developed based on the open-source code DAFoam [33]-[37]. It should be noted that the RANS solver of DAFoam is nearly identical to OpenFOAM [38] and OpenFOAM's SimpleFOAM solver is used in all the test cases of this article. The secondary development is quite convenient thanks to the easily-extensible code structure of DAFoam and the flexibility of the auto-differentiation package CodiPack [39].

## 2.2. Symbolic regression

Like the ANN, RF, and other widely used models, we need a dataset to train an SR model (an analytical expression). The output label of the SR model is the deviation of the optimum correction term found by field inversion from its prior value ($\boldsymbol{\beta}^{opt} - 1$). The input of the model i.e., the local physical features. should be calculated using the flow variables corresponding to $\boldsymbol{\beta}^{opt}$ ($\boldsymbol{w}(\boldsymbol{\beta}^{opt})$). The selection of the input features partly represents the modeler's physical understanding of the correction term $\boldsymbol{\beta}^{opt}$ and may impact the performance of the trained model. In this paper, the following six nondimensional features are chosen as input [1][4][40]:

$$x_0 = tr(\widehat{\boldsymbol{S}}^2) = \lambda_1, x_1 = tr(\widehat{\boldsymbol{\Omega}}^2) = \lambda_2, x_2 = tr(\widehat{\boldsymbol{\Omega}}^2 \cdot \widehat{\boldsymbol{S}}^2) = \lambda_5, x_3 = |\widehat{\boldsymbol{\Omega}}|, x_4 = \frac{|\boldsymbol{\Omega}|d^2}{\nu} = Re_\Omega, x_5 = P/\epsilon \tag{16}$$

$\widehat{\boldsymbol{S}}, \widehat{\boldsymbol{\Omega}}$ are nondimensional strain tensor and rotation tensor respectively:

$$\widehat{\boldsymbol{S}} = \frac{k}{\epsilon}\boldsymbol{S} = \frac{k}{\epsilon}\frac{1}{2}\left(\frac{\partial u_i}{\partial x_j} + \frac{\partial u_j}{\partial x_i}\right)\boldsymbol{e}_i \boldsymbol{e}_j, \widehat{\boldsymbol{\Omega}} = \frac{k}{\epsilon}\boldsymbol{\Omega} = \frac{k}{\epsilon}\frac{1}{2}\left(\frac{\partial u_i}{\partial x_j} - \frac{\partial u_j}{\partial x_i}\right)\boldsymbol{e}_i \boldsymbol{e}_j \tag{17}$$

$|\boldsymbol{\Omega}|$ is the norm of rotation tensor:

$$|\boldsymbol{\Omega}| = \sqrt{\Omega_{ij}\Omega_{ij}}, \Omega_{ij} = \frac{1}{2}(\frac{\partial u_i}{\partial x_j} - \frac{\partial u_j}{\partial x_i}) \tag{18}$$

$d$ is the distance to the nearest wall. $P$ is the production term of the turbulent kinetic energy. The definition can be found in equation (2). $\epsilon$ in equation (16) and (17) is defined as:

$$\epsilon = \beta^* \omega k, \beta^* = 0.09 \tag{19}$$

The first three features in equation (16) are based on Pope's tensor representation theory of Reynolds stress. There are 5 independent tensor invariants for tensor $\hat{S}$ and $\hat{\Omega}$ for 3-D flows, they are named $\lambda_i, i \in \{1,2,3,4,5\}$ by Pope:

$$\lambda_1 = tr(\hat{S}^2), \lambda_2 = tr(\hat{\Omega}^2), \lambda_3 = tr(\hat{S}^3), \lambda_4 = tr(\hat{\Omega}^2 \hat{S}), \lambda_5 = tr(\hat{\Omega}^2 \hat{S}^2) \tag{20}$$

$\lambda_2, \lambda_4$ are identically zero in 2-D flows. $x_0 = \lambda_1, x_1 = \lambda_2$, and $x_2 = \lambda_5$ are the three non-zero invariants among them in 2D flows. Similar features were used in [17] to construct a black-box model for separated flows and achieved good results. The fourth feature is used to measure the rotation of the fluid. The fifth feature is used by [4] to detect off-wall regions where the shear (or the rotation) is quite strong. It is also used in [12] to construct the machine-learning model to predict airfoil-stall. The sixth feature reflects the non-equilibrium characteristic of turbulence, which is used by Rumsey [1] and Li et al.[4]. All the features are chosen based on successful black-box models, analytical corrections, and our experience.

In this paper, the open-source symbolic regression software PySR [21]-[23] is used to discover the analytical relationship between $x_i, i = 0, 1, \cdots, 5$ and $\beta - 1$. PySR allows the user to define custom element functions that are used to build the final analytical expression. It also supports adding various constraints to prevent unreasonable function nesting (i.e., $\exp[\exp(x_1)]$). Element functions that are usually encountered in traditional turbulence modeling are selected for SR:

Table 1. The element function used in the symbolic regression process, $i, j \in \{0,1,2,3,4,5\}$

| *Operator type* | *Operators* |
|---|---|
| *Unary operators* | $\exp(x_i), \tanh(x_i), \dfrac{1}{1+x_i}, \dfrac{1}{x_i}$ |
| *Binary operators* | $x_i + x_j, x_i - x_j, x_i * x_j, \dfrac{x_i}{x_j}, x_i^{x_j}, \min(x_i, x_j), \max(x_i, x_j)$ |

Note that for binary operators, one of their inputs can also be a constant. Constraints are added to prevent $\tanh(\cdot)$ and $\exp(\cdot)$ from nesting in function $\exp(\cdot)$.

In the training process, the sum of the squared error (SSE) is used to measure the accuracy of the expression:

$$SSE = \sum_i [(\beta_i - 1) - y_{i,pred}]^2 \tag{21}$$

The sum is taken over all training samples. $y_{i,pred}$ is the value predicted by the symbolic expression. Note that we train the symbolic expression to fit the difference between $\beta$ and 1.

PySR allows the user to define the complexity of variables, constants, and operators. To make the expression more interpretable, we avoid the expression from containing too many variables by setting the complexity of the variable to 2 and setting the complexity of operators and constants to 1. PySR then computes the complexity of every expression evaluated in the training process by:

$$C(E) = 2N_{var.} + N_{const.} + N_{op.} \tag{22}$$

$N_{var.}$, $N_{op.}$, and $N_{const.}$ are the number of variables, operators, and constants in the expression, $E$ represents the expression being evaluated and $C(E)$ is the complexity of it. The maximum allowable $C(E)$ is set to 16 and any expressions more complex than that will be discarded during training.

PySR uses the following loss function:

$$loss(E) = SSE(E) \cdot \exp(frecency[C(E)]) \tag{23}$$

to keep the expressions in the population having diverse complexities. $frecency[C(E)]$ is the number of expressions that have complexity identical to $E$ generated in a given period divided by a constant. PySR keeps track of the best expression (elite) in every complexity level in the training process and outputs all the elites when the training completes. The final choice is made by the modeler by weighing the complexity, the loss, and the physical interpretability of the expressions.

## 3. Field inversion and the SR model training on the CBFS case

In this section, we introduce field inversion for the multiplicative correction factor $\beta$ and the SR process. We discuss the physical meaning of the selected symbolic expression and make manual revisions based on a priori physical knowledge. The revised expression is integrated into the original SST model, forming the SST-SR model. When applied to the training set (CBFS), the SST-SR model demonstrates superior results compared with the original SST model.

## 3.1. Field inversion on the CBFS case

The computational domain of the CBFS case is shown in Figure 2. Fully-developed boundary layer velocity profile is imposed at the inlet. The maximum velocity at the inlet is $1\ m/s$. The height of the step is $1\ m$ and the Reynolds number based on the height is approximately 13700. 37093 cells are used for the RANS simulation. The height of the first grid layer satisfies $\Delta y^+ < 1$. The grid applied in this case is made available by [2]. [2] used the SimpleFOAM RANS solver, which is also used in this study, and demonstrated that the grid for the CBFS case here is adequately grid-converged.

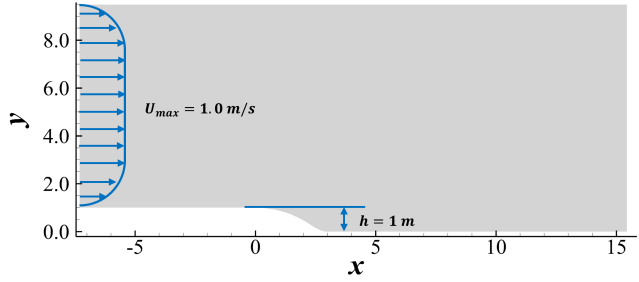

Figure 2. The computational domain of the CBFS case.

The LES simulation was carried out by Bentaleb et al. [41] and was made available by McConkey et al.[2]. In this paper, this LES data is used as high-fidelity data to perform field inversion. For the baseline SST model, the separation zone mainly lies in the blue-shaded area in Figure 3. To correct the predicted separation, 30 sample points are randomly placed in the blue-shaded region and high-fidelity $x$-direction mean velocity data is extracted from them. All the sample points are at least $0.05\ m$ away from the solid wall. The objective function for field inversion is:

$$\min_{\boldsymbol{\beta}} J = 2.0 \sum_i \left(u_i - u_i(\boldsymbol{\beta})\right)^2 + 1.0 \times 10^{-4} \sum_j \left(\beta_j - 1\right)^2 \qquad (24)$$

The $J$ value for the baseline SST model ($\boldsymbol{\beta} = 1$) is 1.15. Note that we choose the velocity data as the target QoI mainly because we assume the underestimated mixing of momentum in the separated shear layer leads to the error of the baseline SST model, and the correction factor $\beta$ is suitable for quantifying this error based on the previous experience [1][4]. Figure 4 shows the convergence history. The optimization converges after 55 iterations with a 90% decrease in the objective function.

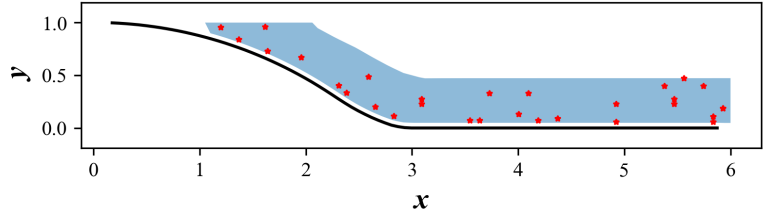

Figure 3. Sample points used in field inversion

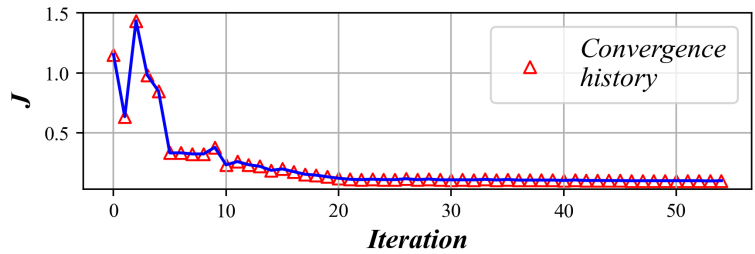

Figure 4. Convergence history of the field inversion on the CBFS case

Figure 5 shows that $\beta$ is increased in the separated shear layer. This is consistent with Rumsey's observation that turbulence activity is often underpredicted in the shear layer by the RANS models [1][4]. The streamline plot in Figure 6 shows graphically that field inversion has significantly suppressed the separation zone. The reattachment point moves upstream by about $1.6\ m$. Figure 7 shows the details of the velocity distribution in the recirculation area. Compared with the baseline result, the velocity profile agrees very well with the LES data.

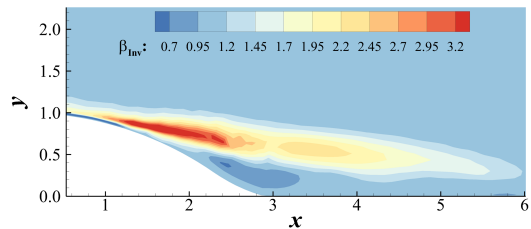

Figure 5. The optimized $\beta$ distribution

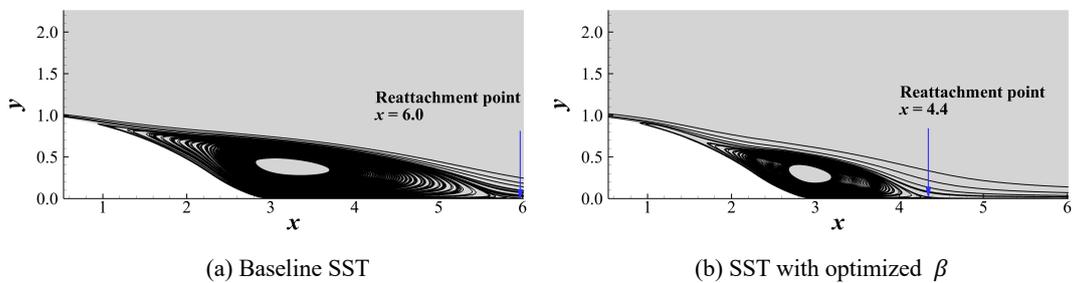

(a) Baseline SST  (b) SST with optimized $\beta$

Figure 6. The separation zone predicted by the baseline SST model and the field inversion

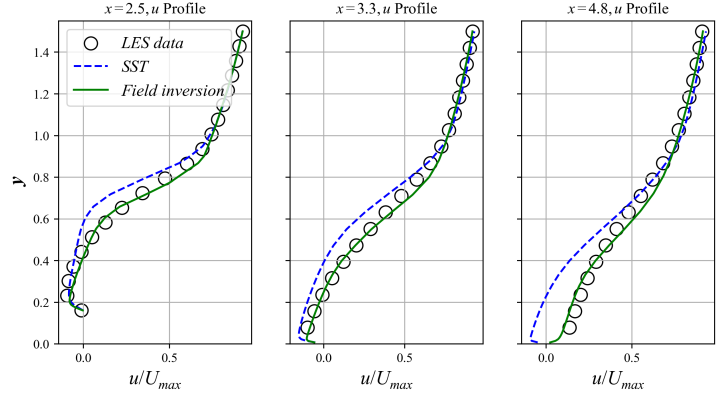

Figure 7. Velocity profiles at different $x$ standpoints in the separation zone.

## 3.2. Symbolic regression using field inversion data on the CBFS case

The data obtained by field inversion in the previous section is used to train the SR model. SR does not require too much training data. To construct the training data, 1000 points are randomly extracted near the separation zone and another 1000 points are selected from the mainstream, as shown in Figure 8. This downsampling strategy helps the SR model learn to distinguish where to activate the correction term and where not to do it. Since the geometry and the separation structure are simple in this case, this simple random-choosing strategy is applicable. For more complicated geometry (i.e., 3D models and 3D separations), more elaborated training set construction and under-sampling technique such as the one used in [15] might be required.

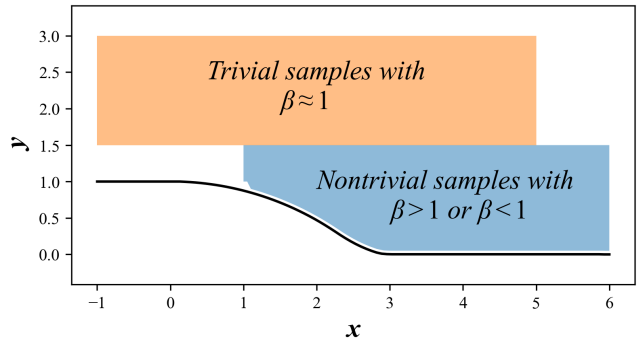

Figure 8. Trivial samples and nontrivial samples

The features used are listed in equation (16). Table 1 shows the element functions used. After 130 generations of evolution that takes about 5 minutes using 10 processors of Intel Xeon Silver 4120R, 15 optimized symbolic expressions with different complexities are generated. All the expressions are listed below in Table 2:

| $C(E)$ | $SSE(E)$ | EQUATION |
|---|---|---|
| 1 | 0.10864 | 0.105013 |
| 4 | 0.032793 | $x_2 * -0.00595966$ |
| 6 | 0.031996 | $min(x_2 * -0.00625924, 2.6413186)$ |
| 7 | 0.031969 | $min(x_2 * -0.006319242, x_5)$ |
| 8 | 0.027482 | $x_2 * (-0.0051232916 - 1/x_4)$ |
| 9 | 0.027481 | $x_2 * (-0.0051232916 - tanh(1/x_4))$ |
| 10 | 0.025439 | $min(1.5593348, (x_1/x_4)x_2)$ |
| 11 | 0.021756 | $min((x_1/x_4)x_2, x_5)$ |
| 12 | 0.019893 | $tanh((x_2 x_1)/x_4)x_5$ |
| 13 | 0.018482 | $tanh(tanh((x_2 x_1)/x_4))x_5$ |
| 14 | 0.017053 | $tanh((x_2 x_1)/x_4)(x_5 - 0.24384463)$ |
| 15 | 0.016371 | $\left(tanh\left(tanh\left(\frac{x_1 x_2}{x_4}\right)\right) x_5\right)^{1.2770432}$ |
| 17 | 0.015853 | $\left(tanh\left(tanh\left(\frac{(4.976726 - x_2)x_3}{x_4}\right)\right) x_5\right)^{1.3573632}$ |
| 19 | 0.015689 | $\left(tanh\left(tanh\left(\frac{(5.226354 - x_2)x_3}{x_4}\right)\right) x_5\right)^{1.3573632} - 0.019771826$ |
| 20 | 0.015293 | $\left(tanh\left(tanh\left(\frac{(4.976726 - x_2)min(x_3, x_0)}{x_4}\right)\right) x_5\right)^{1.3467026}$ |

Table 2. Expressions of different complexities generated by SR on the CBFS case

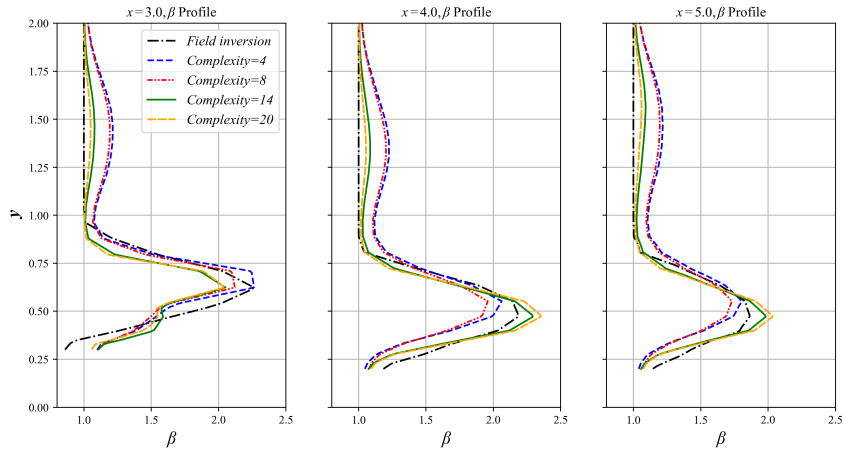

Figure 9. $\beta$ profiles predicted (offline) by expressions with different complexities on multiple standpoints along the separated shear layer

For the expressions with complexity $C(E) \geq 15$, nested $tanh(\cdot)$, power function, and too many other constants emerge, hindering the interpretation of the expression. Furthermore, the decreasing rate of the SSE as complexity grows drops substantially after $C(E) \geq 15$ (from $\frac{d[SSE(E)]}{d[C(E)]} \approx 0.001$ at $C(E) = 14$ to $\frac{d[SSE(E)]}{d[C(E)]} \approx 0.0005$ at $C(E) = 15$). So, increasing complexity cannot decrease the SSE sufficiently after $C(E) \geq 15$. Therefore, we choose the expression with a complexity smaller than 15 and has a minimal SSE i.e., the expression that has a complexity of 14. Figure 9 displays the $\beta$ profiles at various stations adjacent to the separated shear layer. Both the selected expression and the most complex expression demonstrate improved fitting to field inversion results downstream within the separated shear layer and in the region above the separated shear, where $\beta$ is approximately equal to 1. However, the expressions with lower complexities overpredicted $\beta$ above the separated shear layer.

The chosen expression (complexity $= 14$) can be written as:

$$\beta_{SR} = \chi_{SR} + 1 = 1 + \tanh\left(\frac{\lambda_2 \lambda_5}{Re_\Omega}\right)\left(\frac{P}{\epsilon} - 0.244\right) \tag{25}$$

Equation (25) can be interpreted as follows. We first analyze the physical characteristic of the first parenthesis in equation (25). In 2-dimensional incompressible flow, $\lambda_2$ and $\lambda_5$ can be written as:

$$\lambda_2 = tr(\widehat{\boldsymbol{\Omega}}^2) = -|\widehat{\boldsymbol{\Omega}}|^2, \lambda_5 = tr(\widehat{\boldsymbol{\Omega}}^2 \cdot \widehat{\boldsymbol{S}}^2) = -\frac{1}{2}|\widehat{\boldsymbol{\Omega}}|^2|\widehat{\boldsymbol{S}}|^2 \tag{26}$$

Using the condition that $\widehat{\boldsymbol{\Omega}}$ is asymmetric, $\widehat{\boldsymbol{S}}$ is symmetric, and $\nabla \cdot \boldsymbol{u} = 0$, equation (26) can be proved. In the viscous sublayer, applying boundary layer theory and equation (17), we have:

$$|\widehat{\boldsymbol{\Omega}}| = |\widehat{\boldsymbol{S}}| = \frac{1}{\sqrt{2}}\left|\frac{\partial u}{\partial y}\right|\frac{k}{\epsilon} \tag{27}$$

Consequently:

$$\lambda_2 \lambda_5 = \frac{1}{16}\left|\frac{\partial u}{\partial y}\right|^6 \left(\frac{k}{\epsilon}\right)^6 \tag{28}$$

On the other hand, combining equation (16), (17), (18), and (30), $Re_\Omega$ can be expressed as follows in the viscous sublayer:

$$Re_\Omega = \frac{|\Omega|d^2}{\nu} = \frac{1}{\sqrt{2}}\left|\frac{\partial u}{\partial y}\right|\frac{d^2}{\nu} \tag{29}$$

Using equation (28)~(29), and inserting the definition of $\epsilon$ in equation (19), $\frac{\lambda_2 \lambda_5}{Re_\Omega}$ can be expanded as:

$$\frac{\lambda_2 \lambda_5}{Re_\Omega} \propto \frac{\nu}{d^2 \omega^6}\left|\frac{\partial u}{\partial y}\right|^5 \tag{30}$$

Furthermore, $\omega$ can be expressed as follows near the wall [42]:

$$\omega = \alpha \nu / d^2 \tag{31}$$

$\alpha$ is a constant. Inserting (31) to (30), we have:

$$\frac{\lambda_2 \lambda_5}{Re_\Omega} \propto \left|\frac{\partial u}{\partial y}\right|^5 \frac{\nu}{d^2 \omega^6} \propto \left(\left|\frac{\partial u}{\partial y}\right| \frac{d^2}{\nu}\right)^5 \tag{32}$$

Equation (32) shows that $\lambda_2 \lambda_5 / Re_\Omega$ tends to zero rapidly near the wall (as $d$ tends to zero near the wall), making $\chi_{SR}$ approach zero ($\sim O(d^{10})$) in the viscous sublayer.

In mainstream away from the boundary layer, we define:

$$T_t = k/\epsilon \tag{33}$$

The physical meaning of $T_t$ is the time scale of turbulence. If we assume the time scale of the mean flow can be expressed as:

$$\frac{1}{|\Omega|}, \frac{1}{|S|} \approx T_m \tag{34}$$

then $\lambda_2 \lambda_5$ can be rewritten using equations (17) and (26):

$$\lambda_2 \lambda_5 \sim \left(\frac{T_t}{T_m}\right)^6 \tag{35}$$

and $Re_\Omega$ can also be rewritten:

$$Re_\Omega = \frac{d^2}{T_m \nu} \tag{36}$$

Consequently, $\lambda_2 \lambda_5 / Re_\Omega$ can be expressed as follows by combining equation (35)~(36):

$$\frac{\lambda_2 \lambda_5}{Re_\Omega} \approx \left(\frac{T_t}{T_m}\right)^5 \frac{T_t \nu}{d^2} \tag{37}$$

In typical CFD computations with $k$ and $\omega$ given at the inlet, $T_t$ remains constant in the mean flow away from the wall. In the absence of strain, expansion, or compression in the mean flow, $T_m \approx 1/|S|$ approaches infinity, and with large $d$, $1/d$ tends towards zero. Consequently, equation (28) generally tends to zero in the mainstream where fluid deformation is weak (in the CBFS case, it is about $1 \times 10^{-3}$ in the mainstream). In separated shear layers, the strain is large, and $T_m \approx 1/|S|$ is nearly comparable to $T_t$. Additionally, the shear layer is close to the wall, making $1/d^2$ large. Therefore, $\lambda_2 \lambda_5 / Re_\Omega$ is approximately $O(1)$ in the separated shear layer (it is about 1~10 in the CBFS case). After applying the $\tanh(\cdot)$ operator, it remains $O(1)$ but does not exceed 1. These arguments suggest that $\tanh\left(\frac{\lambda_2 \lambda_5}{Re_\Omega}\right)$ acts as an activation function, turning the correction term on in regions with strong strain and not far from the wall, and off in the mainstream where no deformation occurs.

We now examine the second parenthesis in equation (24). Given the first parenthesis acts as an activation function with values between 0 and 1, this term indicates that when $\tanh(\frac{\lambda_2 \lambda_5}{Re_\Omega})$ is activated (~1), the correction term's magnitude depends linearly on $P/\epsilon$, representing turbulence non-equilibrium. In separated shear layers, non-equilibrium turbulence prevails [1][4]. Larger $P/\epsilon$ results in a greater $\beta$, causing stronger $\omega$ destruction and weaker $k$ dissipation, thereby increasing turbulence activity (higher $k$). This elevation in turbulent kinetic energy leads to a rise in $\mu_T$, enhancing momentum exchange between the separated shear layer and the mainstream and promoting reattachment by increasing the momentum of the shear layer. The physical process above is illustrated in Figure 10.

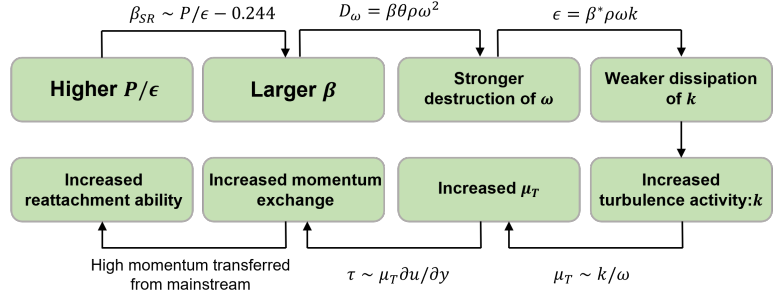

Figure 10. The physical mechanism of increasing the reattachment ability of the separated shear layer implied by the correction term.

As shown in Figure 10, the correction term promotes turbulent activity in non-equilibrium turbulence. Consequently, the correction term is physically consistent with Rumsey's observation [1], which states that the RANS models often underpredict the turbulence activity in non-equilibrium turbulence (i.e., $\frac{P}{\epsilon} > 1$).

The discussion above analyzed every term in equation (25) physically, demonstrating good interpretability of the model (expression) generated by SR. However, the expression in equation (25) is not perfect. The activation term $\tanh\left(\frac{\lambda_2 \lambda_5}{Re_\Omega}\right)$ is about 1 near the edge of the attached boundary layer, which is not preferable. Furthermore, in the region of expansion or compression (such as the sudden expansion after the backward-facing step), $|S| \neq 0$, making $\tanh\left(\frac{\lambda_2 \lambda_5}{Re_\Omega}\right) \sim O(1)$. To prevent these wrongly activated correction, two other switches are added to the final expression:

$$\beta_{SR} = \chi_{SR} s_{\lambda_5} s_I + 1.0 \tag{38}$$

$$s_{\lambda_5} = \frac{1}{2}\tanh[C_{\lambda_5,1}(\lambda_5 - C_{\lambda_5,2})] + \frac{1}{2}, s_I = \frac{1}{2}\tanh[C_{I,1}(I - C_{I,2})] + \frac{1}{2}$$

$$I = k/|\mathbf{u}|^2$$

$$C_{\lambda_5,1} = -5.0, C_{\lambda_5,2} = -27.0, C_{I,1} = 800, C_{I,2} = 0.007$$

An illustration of the function $\frac{1}{2}\tanh[C_1(q - C_2)] + \frac{1}{2}$ is shown in Figure 11. The curve jumps from 0 to 1 at $q = C_2$, and $C_1$ is proportional to the tangent at $q = C_2$. The larger $|C_1|$ is, the more abrupt the jump. Consequently, $\frac{1}{2}\tanh[C_1(q - C_2)] + \frac{1}{2}$ can be viewed as an activation function that switches between 0 and 1 as $q$ varies. By multiplying $s_{\lambda_5}$ to $\chi_{SR}$, it means that the correction will be activated only if $|\lambda_5|$ is larger than $|C_2|$. It is based on our observation that $|\lambda_5|$ is significantly higher in the separated shear layer than in the boundary layer. On the other hand, multiplying $s_I$ to $\chi_{SR}$ can deactivate the correction term where the turbulence intensity ($I$) is smaller than $C_{I,2}$. The physical meaning of $s_I$ is that $\chi_{SR}$ should not be turned on in the region where turbulence activity is very weak (i.e., the expansion of mainstream). All the constants are calibrated on the CBFS case.

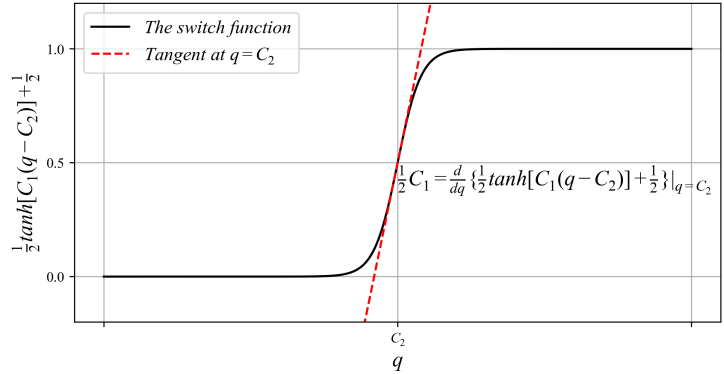

Figure 11. The illustration of the switch function $s_q$.

## 3.3. CFD-coupled prediction using the SST-SR model on the CBFS case

We integrate the modified SR expression, equation (38), into OpenFOAM by simply typing the expression into the source code of the original SST model. We call this modified SST model the SST-SR model. Then, the SimpleFOAM solver coupled with the SST-SR model is used to predict the flow in the CBFS case. The convergence levels of the SST-SR model and the SST model are approximately the same (the residual drops to $1 \times 10^{-11}$ after 5000 iterations), with the SST model converging faster. On the other hand, it takes the SST-SR model about 33% more time to complete 5000 iterations compared with the SST model (24 cores of Intel Xeon 4210R CPU are used). This is caused by extra gradient evaluation and tensor algebra calculation required by the features used by the correction term $\beta_{SR}$. Since we did not try to optimize the performance of our code, the space for speed-up still exists.

The contour of the predicted $\beta_{SR}$ is shown in Figure 13(a). It is quite similar to the $\beta$ field obtained by field inversion, which is shown in Figure 5. The predicted separation zone in Figure 13(b) is nearly identical to the results given by the field inversion in Figure 6(b) and is substantially improved compared with the results given by the original SST model. The difference between the reattachment point predicted by SST-SR and field inversion is only $0.1\ m$. The velocity profile shown in Figure 14 shows that the SST-SR model outperforms the SST model substantially, agreeing well with the LES data.

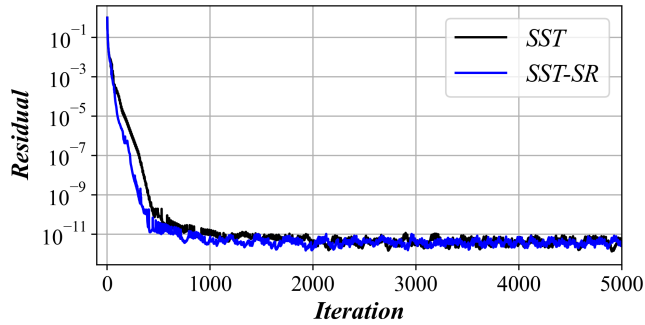

Figure 12. Residual plots in the converging history of the SST model and the SST-SR model

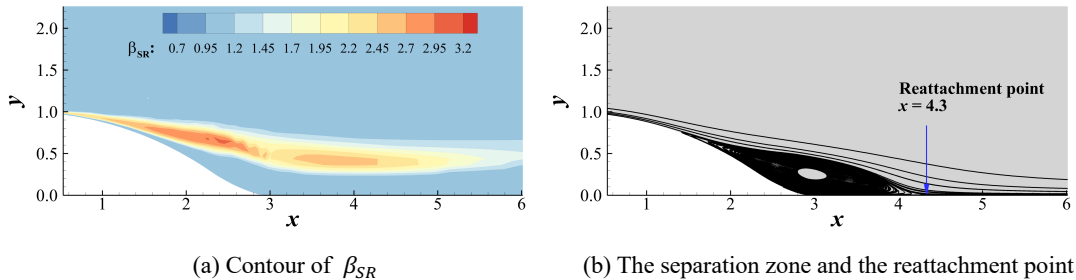

(a) Contour of $\beta_{SR}$      (b) The separation zone and the reattachment point

Figure 13. The predicted $\beta_{SR}$ and flow separation by the SST-SR model.

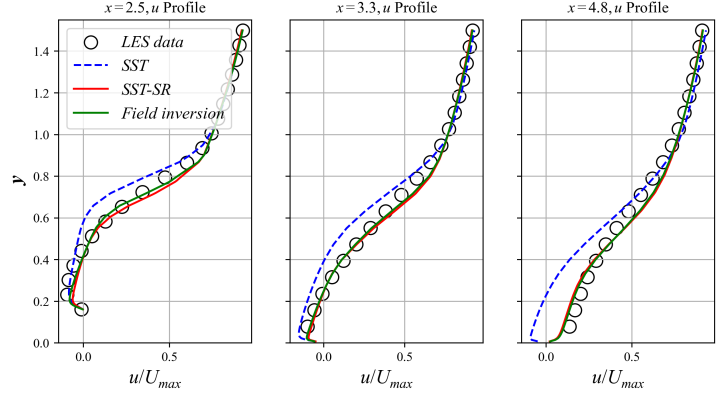

Figure 14. Comparing the velocity profiles predicted by different methods with the LES data.

The result above indicates that the modified SR expression gives satisfactory results when it performs CFD-coupled prediction, even though the expression itself is rather simple.

## 4. Distinct test cases of the SST-SR model

In this section, the SST-SR model is tested on various test cases completely different from the training set. Strong generalizability of the SST-SR model is shown in these tests.

## 4.1. 2D-bump with various height

The SST-SR model is applied to 2-dimensional parametric bumps with maximum heights of $42\ mm$, $38\ mm$, $31\ mm$, and $26\ mm$ [43]. The Reynolds number of these test cases ranges from 17,240 to 27,850. Figure 15 illustrates the bump geometries. Several differences exist between this test case and the training set (the CBFS case):

1. The geometry differs significantly. In the CBFS case, the flow undergoes expansion due to the step, while in the 2D-bump case, it experiences both expansion (deceleration) and compression (acceleration).
2. The separation zone characteristics vary. The CBFS case exhibits a large separation covering most of the step and extending downstream, while in the 2D-bump case, a low bump height results in a small separation zone, with the flow essentially remaining attached to the wall.
3. The flow variable numerical values are distinct. In the CBFS case, the maximum inlet velocity and step height are both normalized to 1, whereas in the 2D-bump case, they are not normalized, with $U_{in} \approx 18\ m/s$ and $h \in [26\ mm, 42\ mm]$. As a result, the $|S|$ in the CBFS case around the separated shear layer is about 3~6, whereas in the 2D-bump case, $|S|$ is about 1000 in the separated shear layer.

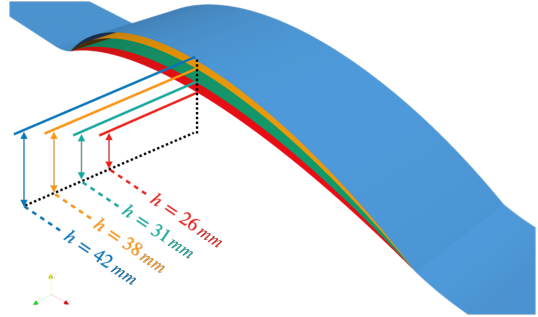

Figure 15. 2D parametric bumps with different maximum height

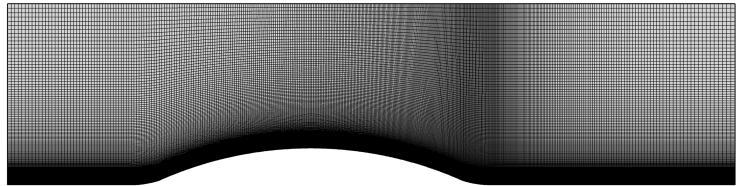

Figure 16. The mesh of 2D bump case, $h = 31\ mm$

The mesh used is shown in Figure 16. It contains 72100 cells. The meshes for all bump cases are taken from [2]. [2] used the same RANS solver (SimpleFOAM) to study the mesh convergence and found that the mesh applied for 2D-bump cases is grid-converged. The inlet velocity profile is specified based on the boundary layer thickness in the LES computation [43]. The zero-gradient boundary condition is specified at the upper boundary and the right boundary. The bottom boundary is treated as a solid wall. Field inversion is performed on all 2D-bump cases using the available LES data in [2]. For each case, 30 random sample points are chosen from the recirculation zone as high-fidelity data. For field inversion, the prediction error for velocity decreases by over 90%. The velocity field obtained by field inversion serves as a reference for qualitatively evaluating the SST-SR model's accuracy. Results for $h = 42\ mm$ and $h = 31\ mm$ are discussed here, while computations for $h = 38\ mm$ and $h = 26\ mm$ are presented in the appendix for clarity.

In the $h = 42\ mm$ case, Figure 17 demonstrates that the recirculation zones given by field inversion and the SST-SR model both align with the LES data well, while the SST model incorrectly predicts a delayed reattachment. With the reattachment point provided by LES at approximately $x_{reattach} = 0.340$ [43], the SST-SR model reduces the reattachment point prediction error by 80.0%. As depicted in Figure 18, $\beta_{SR}$ increases to around 2 in the region above the separated zone, which corresponds well with the observation of non-equilibrium turbulence prevailing over the shear layer in [43]. Figure 19 presents the velocity profiles in the recirculation zone from different methods, revealing that the SST-SR model significantly outperforms the SST model and more closely matches the LES data. The TKE profile is shown in Figure 20. The result given by the SST-SR model agrees well with the LES data, while the SST model gives a peak TKE deviating from the LES data.

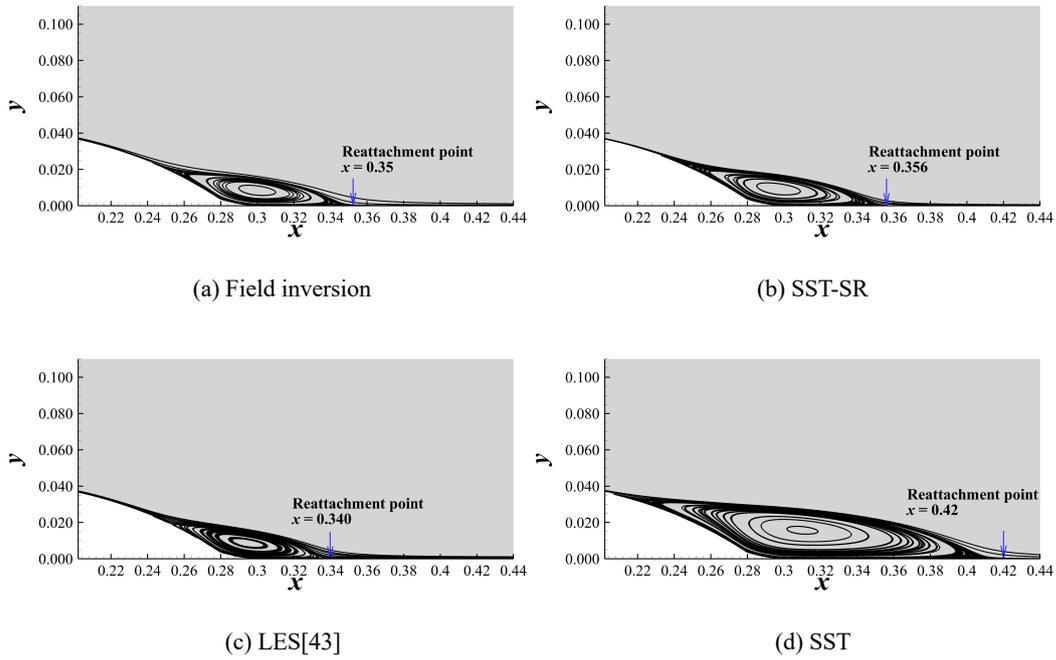

Figure 17. Separation zone given by different methods, $h = 42\ mm$.

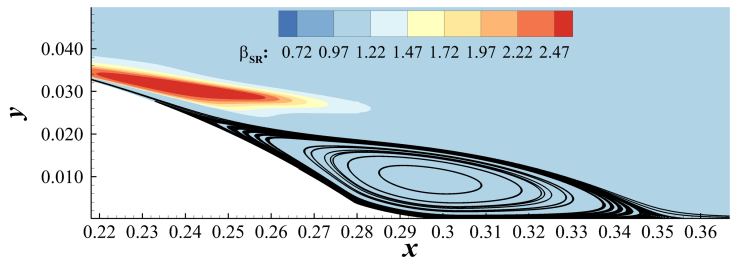

Figure 18. The $\beta_{SR}$ distribution in the $h = 42\ mm$ case predicted by the SST-SR model

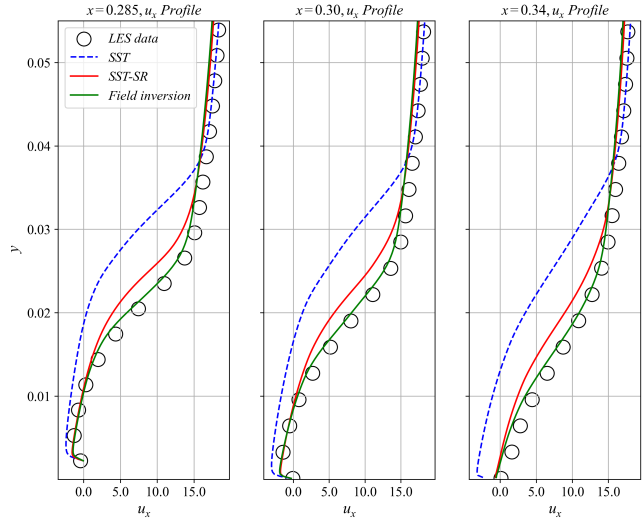

Figure 19. The velocity profile predicted by different methods, $h = 42\ mm$.

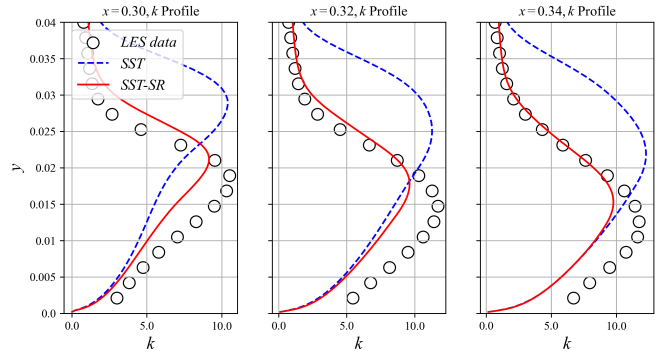

Figure 20. The TKE profile predicted by the SST and the SST-SR model, $h = 42\ mm$

In the $h = 31\ mm$ case, the SST-SR model, similar to the $h = 42\ mm$ case, predicts a smaller recirculation zone than the SST model, better aligning with the field inversion results and the LES data as shown in Figure 20. Based on the LES results from [43] ($x_{reattach} = 0.303\ m$), the reattachment point error is reduced by 95%. Figure 21 reveals that the region with high $\beta_{SR}$ is over the separated shear layer, mirroring the $h = 42\ mm$ case. The velocity profile in Figure 22 showcases the SST-SR model's enhanced ability to predict the recirculation zone compared to the SST model. Figure 24 also shows that the TKE profile predicted by the SST-SR model matches the LES data better compared with the SST model.

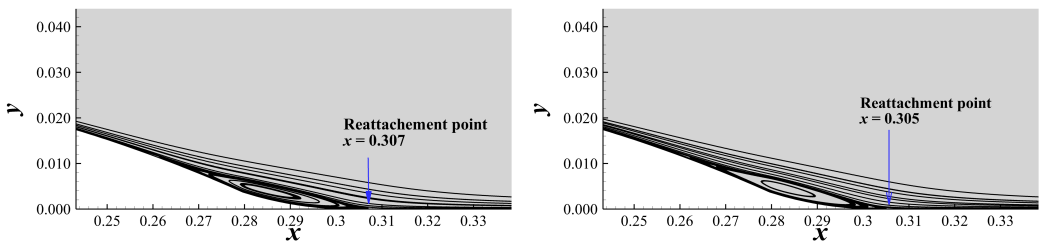

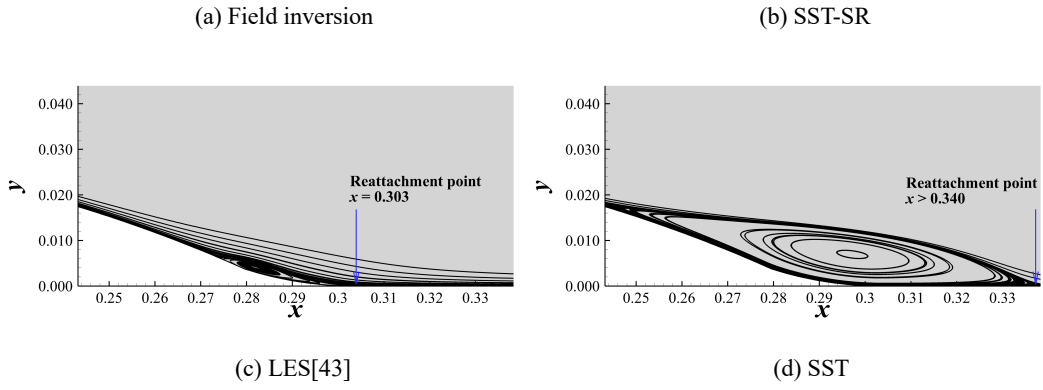

(a) Field inversion  (b) SST-SR

(c) LES[43]  (d) SST

Figure 21. Separation zone given by different methods, $h = 31\ mm$.

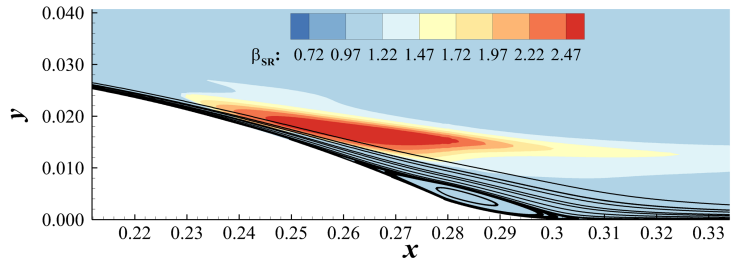

Figure 22. The $\beta_{SR}$ distribution in the $h = 31\ mm$ case predicted by the SST-SR model

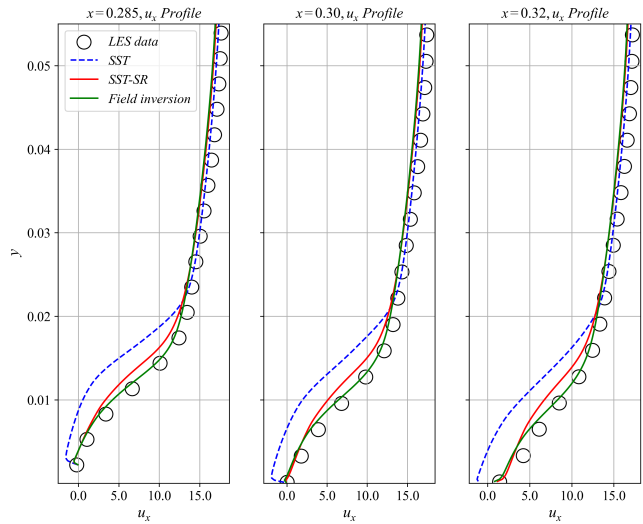

Figure 23. The velocity profile predicted by different methods, $h = 31\ mm$.

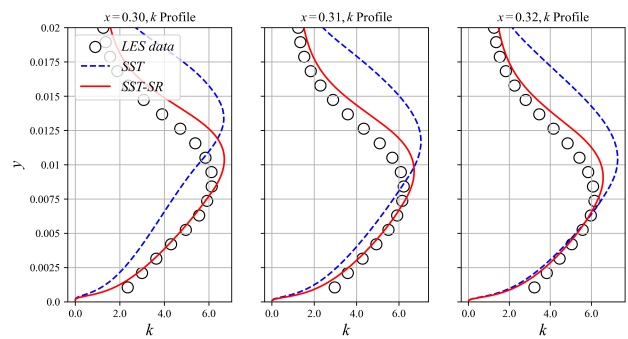

Figure 24. The TKE profile predicted by the SST and the SST-SR model, $h = 31\ mm$

In summary, the 2D-bump test case demonstrates that the SST-SR model can generalize to cases with varying geometry, separation shape, Reynolds number, and numerical values.

## 4.2. Periodic hill

In this section, we evaluate our model on the periodic hill using available DNS data [44], with the Reynolds number based on the hill's height $Re_H = 5600$. In [44], a series of periodic hills are defined with different geometric variables $\alpha$. In this paper, the geometry that has a height of $L_y = 3.036$ and a length of $L_x = 3.828\alpha + 5.142 = 8.228$ is chosen, corresponding to the $\alpha = 0.8$ case in [44]. The mesh (Figure 25) consists of 14,751 cells, and cyclic boundary conditions are applied on the left and right boundaries, while solid-wall conditions are enforced on the upper and lower boundaries. We call this mesh the medium mesh. A coarser mesh consisting of approximately 7000 cells and a finer mesh with about 25000 cells are also made for grid convergence study. The results of this section are mainly from the medium mesh, but we will show that enough grid convergence is achieved by the medium mesh. Figure 26 reveals an increase in $\beta_{SR}$ (1.3~2.0) near the separated shear layer and at the windward side of the second crest due to strong curvature effects. Figure 27 shows that the flow reattaches at $x = 5.2$ according to DNS data, while the SST model predicts reattachment in the middle of the second hill. The SST-SR model suggests a slightly delayed reattachment point at $x = 6.2$.

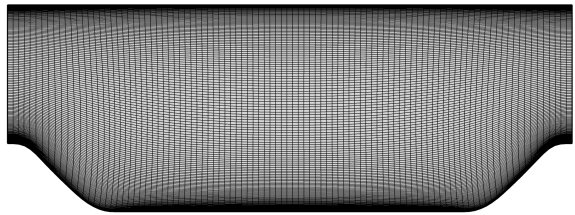

Figure 25. The computational mesh (medium) of the periodic hill case

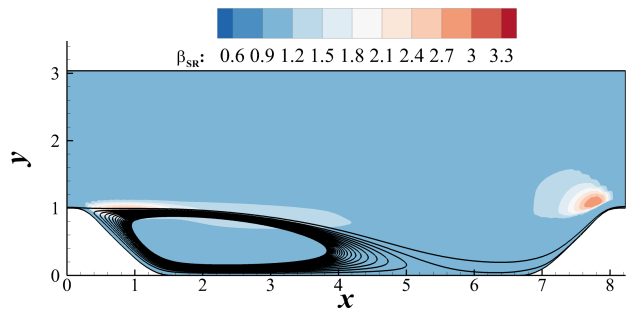

Figure 26. The $\beta_{SR}$ distribution, medium mesh

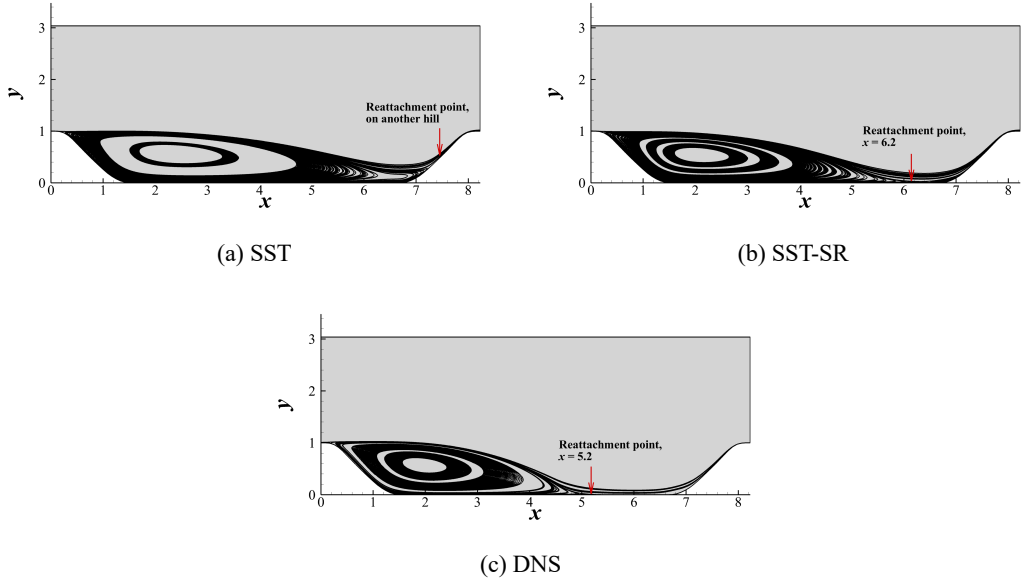

(a) SST

(b) SST-SR

(c) DNS

Figure 27. The separation zone predicted by different methods, medium mesh

Figure 28 presents a scatter plot of all 14,571 data points extracted from cell centers, where the $x$-axis represents the mean $u_x$ from DNS data and the $y$-axis represents the mean $u_x$ from various RANS models. Greater model accuracy is indicated by data points clustering near the 45° line. The SST-SR model demonstrates higher accuracy, with data points aligning more closely to the 45° line. Its mean squared error (MSE) for predicted mean velocity is only 34.6% of the SST model's MSE. Data points in the blue-shaded region represent overpredicted recirculation (true $u_x > 0$, predicted $u_x < 0$). The SST-SR model exhibits fewer data points in this region, indicating a reduced false recirculation zone. The velocity profiles given by the SST model and the SST-SR model on the coarse mesh, medium mesh, and the fine mesh are shown in Figure 29. Results given by different mesh are very close, indicating a good grid convergence is achieved on the medium mesh. On the other hand, the SST-SR model gives a more accurate prediction of velocity compared with the SST model. Figure 30 shows the TKE profiles on the medium mesh, the SST-SR model also outperforms the SST model in the accuracy of TKE.

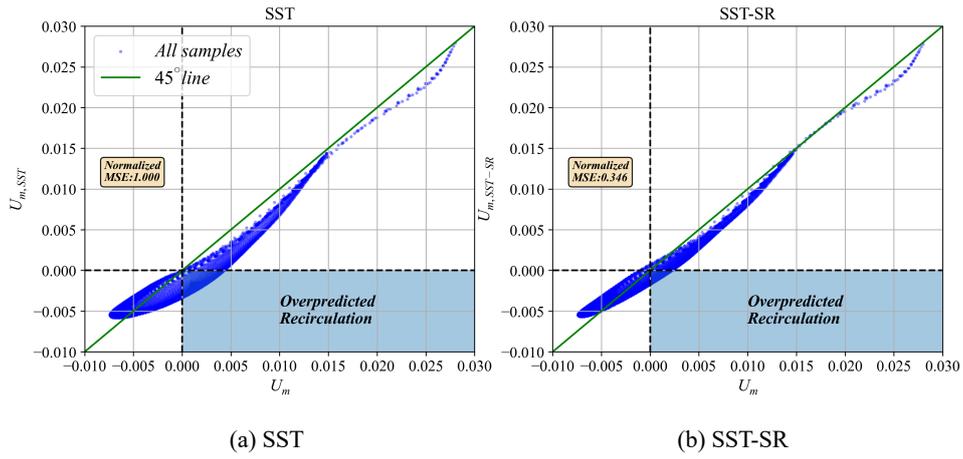

(a) SST  (b) SST-SR

Figure 28. The scattered plot of $u_x$ predicted by DNS and RANS models in every cell, medium mesh

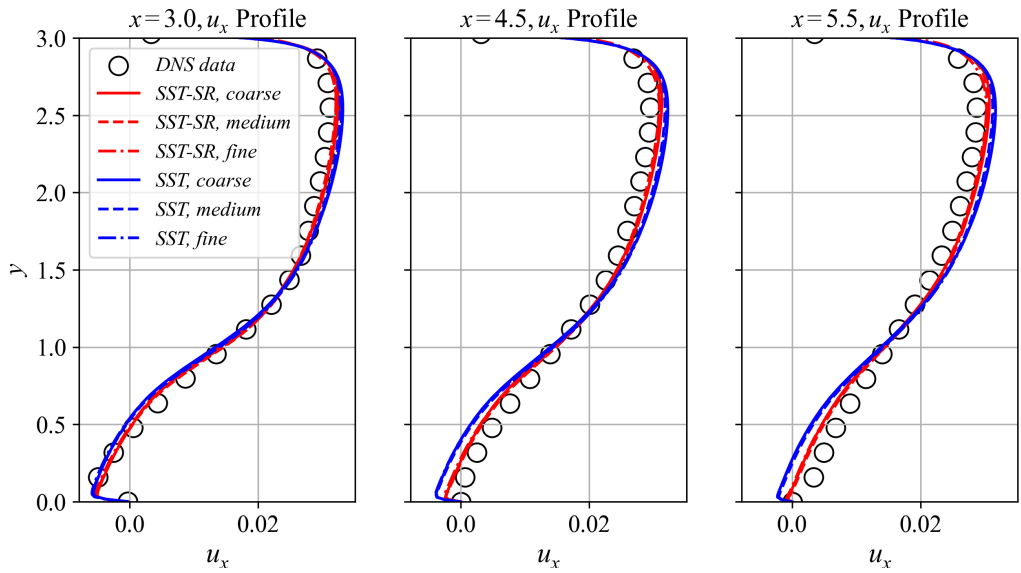

Figure 29. Velocity profiles at different standpoints in the separation zone. The results on coarse mesh, medium

mesh, and fine mesh are compared.

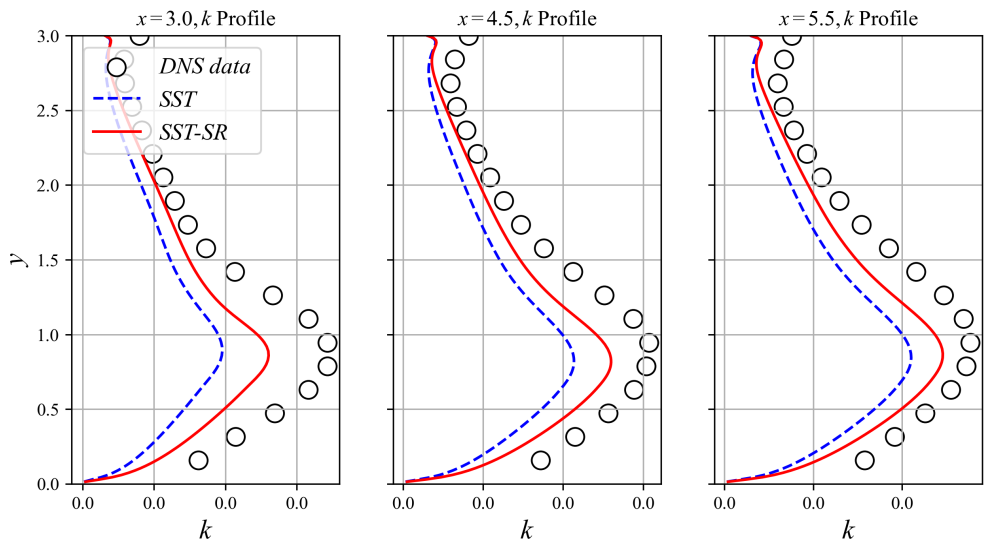

Figure 30. TKE profiles on the medium mesh given by the SST-SR model and the SST model

## 4.3. Ahmed body (3D)

The SST-SR model is applied to the Ahmed body [45] to evaluate its ability to predict 3-dimensional complex separated flows. Often employed as a simplified car model, the Ahmed body helps understand the flow field surrounding an automobile. Figure 31 presents the geometry of the Ahmed body, with a Reynolds number based on body length ($Re_L$) of $2.78 \times 10^6$ and a slant angle ($\phi$) of 25.0 degrees. The computational domain is depicted in Figure 32, utilizing a half-model with a symmetry plane at $y = 0$. No-slip boundaries are established on $z = 0$ and the Ahmed body, while far-field conditions are applied to all other boundaries. The mesh near the Ahmed body is shown in Figure 32, with approximately $3.6 \times 10^6$ hexahedral cells used.

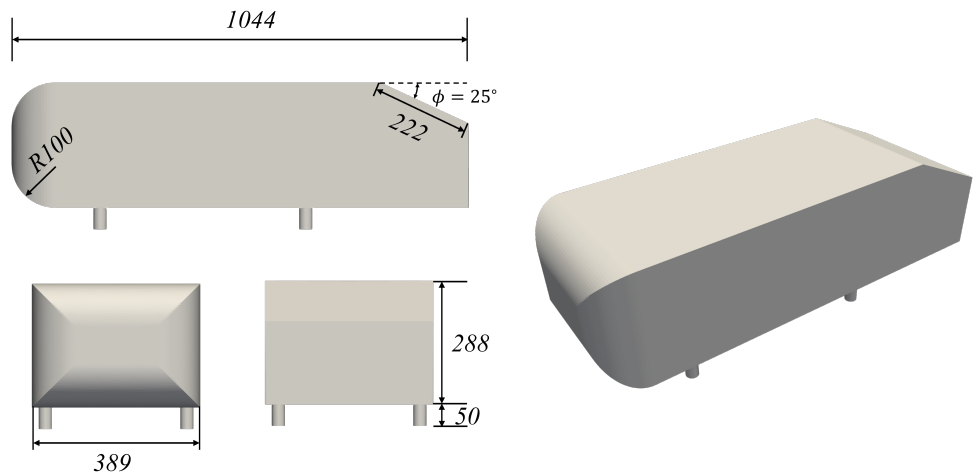

Figure 31. The geometry of the Ahmed body

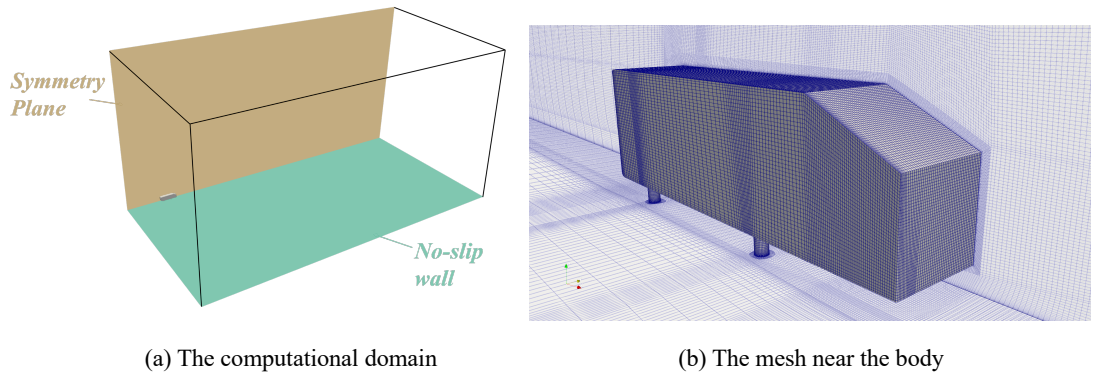

(a) The computational domain          (b) The mesh near the body

Figure 32. The computational domain and the mesh

The findings in [46][47] demonstrated that both the SST model and the DES method predict an extensive separation zone originating from the slant's beginning and a broad wake defect. Our calculations using the SST model, as depicted in Figure 33(a), also reveal a complete flow separation on the slant, resulting in a massive recirculation zone nearly as high as the Ahmed body. However, the experiment [45] indicated that the flow remains attached to the slant in the symmetry plane, yielding a comparatively narrow wake defect. Figure 33(c) presents the PIV results (near-wall data is unavailable). Note that Figure 33(c) is generated by [47] based on the data in [45]. Figure 33(b) displays the wake predicted by the SST-SR model, which features a substantially reduced separation region compared to the SST model, aligning better with experimental data. To study if the solution (with about 3.6 million cells) is grid-converged, we increase the number of cells to obtain a finer mesh (with about 6.3 million cells) and apply the SST-SR model to it. Figure 34 shows that the velocity profiles obtained by the SST-SR model on the original mesh and the finer mesh are very close, demonstrating good grid convergence of the solution on the original mesh. Figure 34 also shows that the SST-SR model outperforms the SST model on both meshes, obtaining a smaller wake defect that matches better with the experimental data.

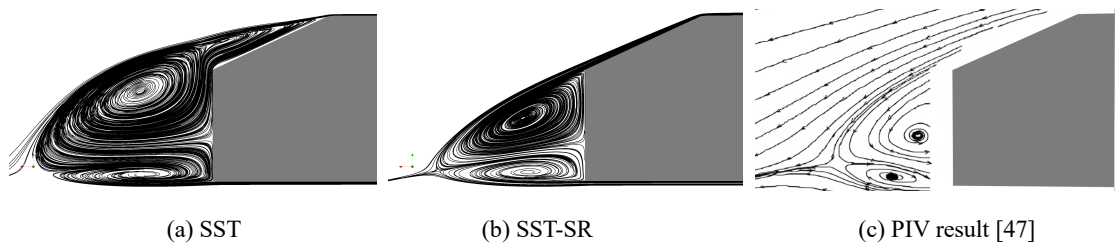

(a) SST                    (b) SST-SR                    (c) PIV result [47]

Figure 33. Streamline plot near the slant and the wake, on the symmetry plane. Note that the streamline plot made in [47] is based on the data in [45]

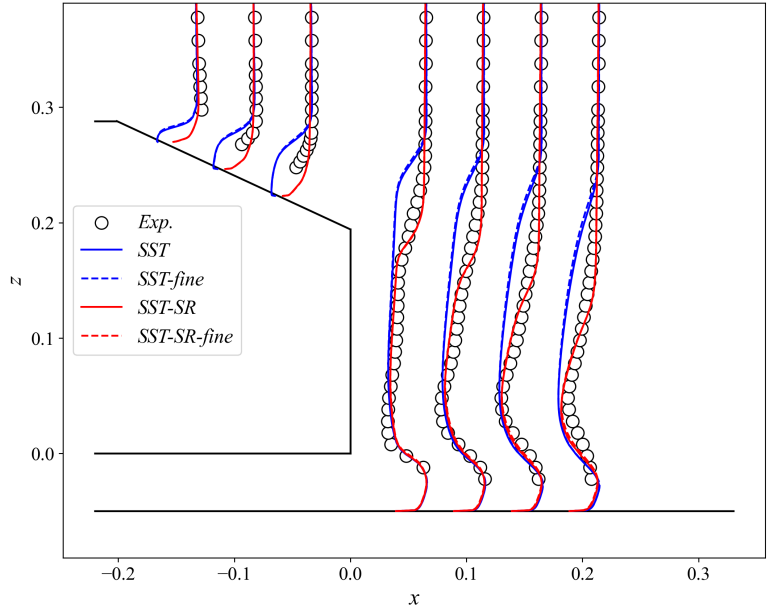

Figure 34. Velocity profiles on the symmetry plane

Figure 35 visualizes the 3D structure of the flow field by displaying the vorticity magnitude $|\Omega|$ on multiple slices in the wake (data has been mirrored against the symmetry plane). Two powerful vortices, often referred to as C-pillar vortices in the automotive industry, are generated by the slant's edge. In the SST model's results, the $|\Omega|$ contour expands along the $y$ and $z$ axes, signifying a wider and taller separation zone. In contrast, the SST-SR model presents a more concentrated distribution of $|\Omega|$, indicative of a smaller recirculation in the wake.

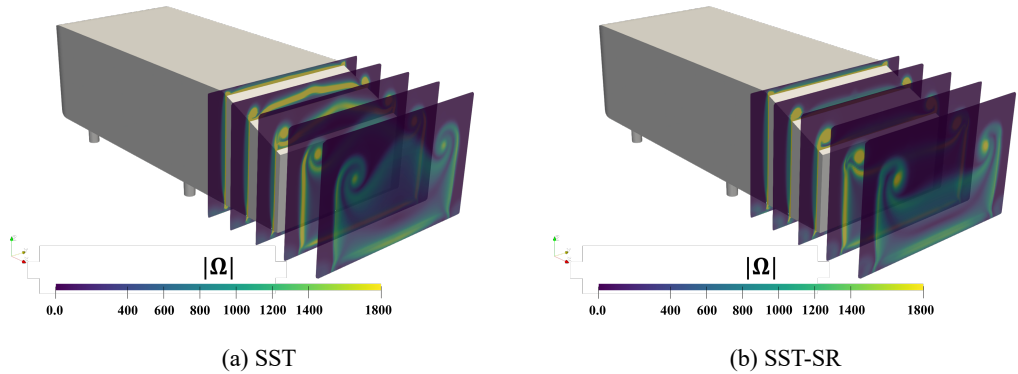

(a) SST  (b) SST-SR

Figure 35. Vorticity magnitude slices near the wake

The correction term $\beta_{SR}$ in the SST-SR model is primarily activated on the slant, where an adverse pressure gradient dominates. Figure 36 depicts $\beta_{SR}$ distribution in multiple $y = const.$ planes intersecting the slant. In this region, $\beta_{SR}$ increases to 2~3, reducing turbulence dissipation and promoting momentum mixing near the slant, allowing the flow to remain attached in most areas.

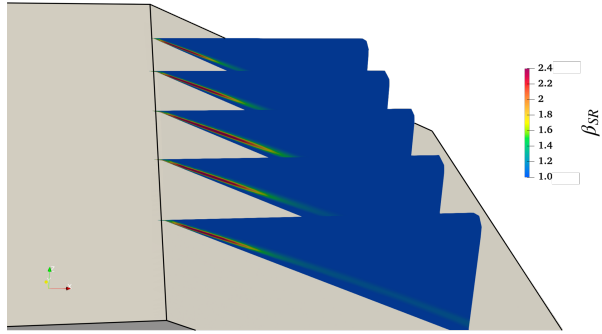

Figure 36. $\beta_{SR}$ is increased to 2~3 at the beginning of the slant.

In summary, the SST-SR model surpasses the SST model in the Ahmed-body case, illustrating its capability to predict complex 3D separation structures. This also highlights the generalization power of the data-driven turbulence model provided by SR.

## 4.4. Turbulent boundary layer on a flat plate

The baseline SST model, already validated for benchmark flows like turbulent boundary layers, should not be negatively impacted by our correction. However, as shown in [19], typical ANN correction models derived from standard FIML procedures to improve accuracy in separated flows often underperform in simple attached flows such as zero-pressure-gradient (ZPG) turbulent boundary layers. We assess the data-driven SST-SR model on a ZPG turbulent boundary layer over a flat plate with $Re_L = 1.0 \times 10^7$, using a rectangular grid and a $\Delta y^+$ of 0.05 for the first grid layer to accurately resolve the viscous sublayer. About $2 \times 10^5$ cells are used for the simulation. Figure 37 illustrates the computational domain and the boundary condition. Figure 35 displays the velocity profiles at $Re_x = 0.25 \times 10^7$ and $Re_x = 0.5 \times 10^7$, with the SST and SST-SR models yielding nearly identical profiles and effectively resolving the viscous sublayer and log layer.

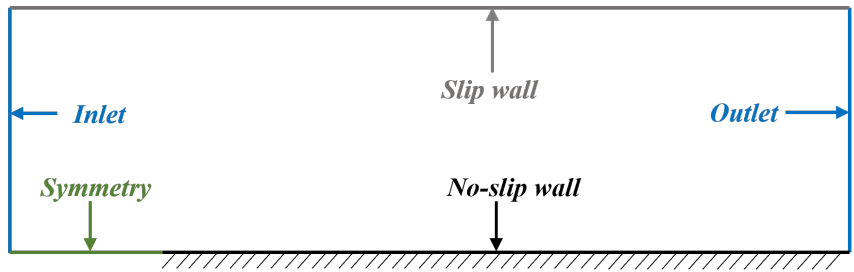

Figure 37 The computational domain of the ZPG flat plate.

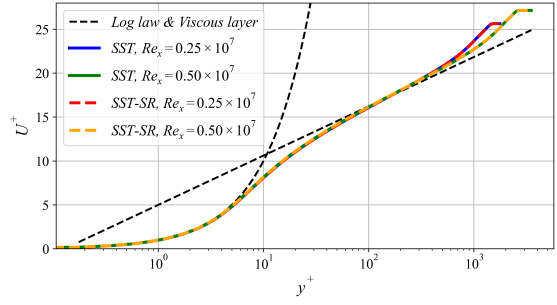

Figure 38. Velocity profile at $Re_x = 0.5 \times 10^7$

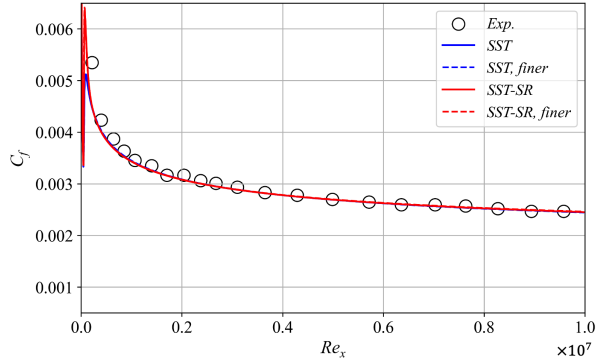

Figure 39. $C_f$ distribution along the flat plate.

The $C_f$ distribution plot in Figure 39 reveals that the SST-SR model's results align well with experimental data [48]. To check if the grid used here is fine enough to achieve grid convergence, a finer mesh with $\Delta y^+ \approx 0.025$ is generated. The number of cells is approximately $4 \times 10^5$ (refined in both directions compared with the baseline mesh). The result on the finer mesh is also plotted in Figure 39. It shows that the difference between the finer mesh's result and the original result is very small, demonstrating good grid convergence of the solution on the original mesh. The $C_f$ given by the SST-SR model is lightly larger than the SST model near the flat plate's leading edge. This is attributed to the increased $\beta_{SR}$ in a small region at the front of the plate (see Figure 40), which intensifies turbulence activity, as demonstrated in Figure 41. The region with increased $\beta_{SR}$ extends only to about $Re_x = 1 \times 10^5$ and $\beta_{SR}$ remains 1 elsewhere. The adverse pressure gradient depicted in Figure 42 near $x = 0.01\ m$, caused by the leading edge, is a primary reason for the elevated $\beta_{SR}$. Figure 43 illustrates the physical mechanism that causes the adverse pressure gradient at leading edge of the plate.

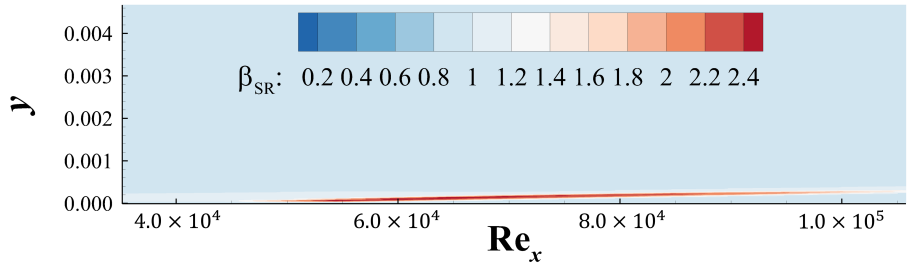

Figure 40. $\beta_{SR}$ distribution at the most front part of the flat plate

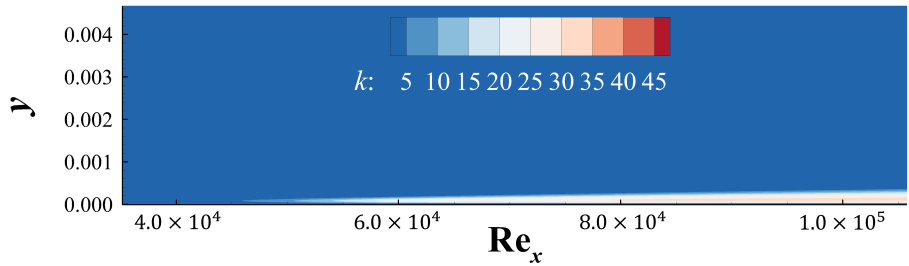

(a) SST

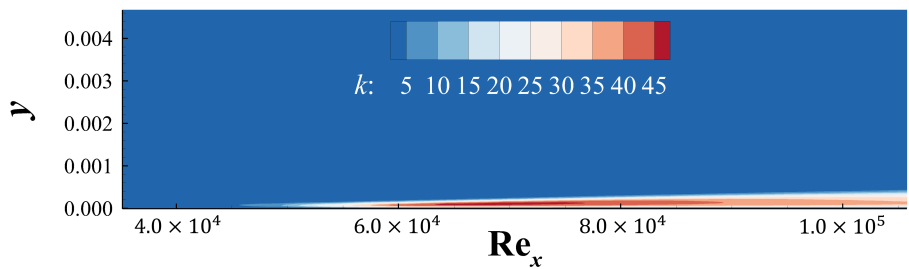

(b) SST-SR

Figure 41. Turbulent kinetic energy ($k$) distribution

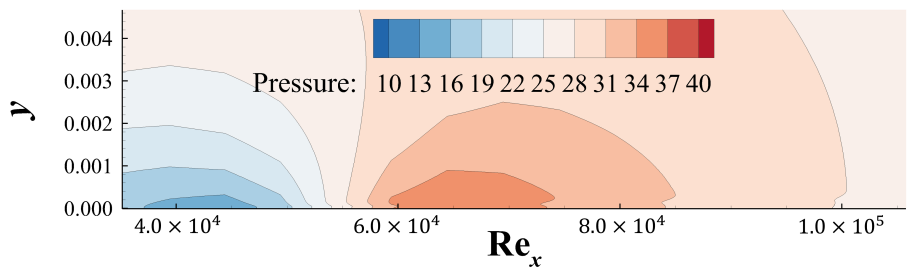

Figure 42. The adverse pressure gradient near the leading edge of the flat plate.

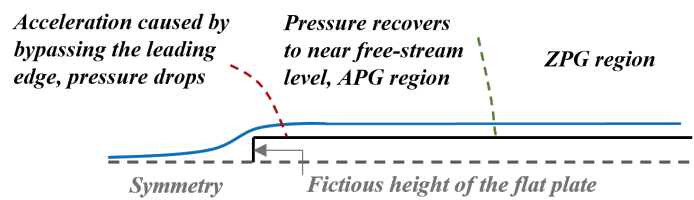

Figure 43. The physical mechanism that forms the region with adverse pressure gradient.

# 5. Conclusions

In this study, we present a generalizable data-driven turbulence model developed through field inversion and symbolic regression. Field inversion is performed on the CBFS case to obtain the optimized $\beta$ distribution. Using the optimized $\beta$ and flow features $\mathbf{w}$, we applied the symbolic regression algorithm to derive a compact, interpretable (as demonstrated in section 2.2) analytical expression for $\beta$ ($\beta_{SR} = \beta_{SR}(\mathbf{w})$). This expression was integrated into OpenFOAM, resulting in the SST-SR model. We then tested the SST-SR model on its training set (the CBFS case) and various distinct cases. The following conclusions can be drawn from the results:

1. The SST-SR model surpasses the SST model in both the 2D bump case and the periodic hill case, demonstrating its ability to generalize to 2D separated flows with characteristics entirely different from the training set.
2. In the 3D Ahmed body case, the SST-SR model produces milder separation, more closely aligning with the PIV data compared to the SST model. This highlights the SST-SR model's capability to predict 3D complex separated flow, which also greatly differs from its training set.
3. The SST-SR model's predictions for friction coefficient and velocity profiles in the turbulent boundary layer case align well with experimental data and theory, showing no signs of weakening the baseline model's ability to predict ZPG attached flows.

In conclusion, the data-driven SST-SR model, generated using symbolic regression, demonstrates a strong generalization ability. To the best of the authors' knowledge, this level of generalizability has not been shown by data-driven turbulence models in the existing literature. The results also reveal the potential of the FISR framework to discover physically interpretable models, suggesting a novel approach for data-driven turbulence modeling that produces more physically-based, generalizable, and portable models.

# Appendix: 2D-bump results of $h = 26\ mm$ and $h = 38\ mm$

    Four 2-D bump cases with different heights are calculated in this study. Two of them are discussed in section 4.1 and another two of them ($h = 38\ mm$ and $h = 26\ mm$) are shown here. The separation zone at $h = 26\ mm$ is shown in Figure 44. The reference separation zone given by field inversion and the LES data is very small, which is very similar to the result of the SST-SR model. However, the SST model gives a significantly larger separation zone. The velocity profile in Figure 45 also shows that the SST-SR model's prediction aligns with the LES data better than the SST model. The TKE profiles given by the SST-SR model are also better, shown in Figure 46. As shown in Figure 47, $\beta_{SR}$ is increased to about 2.5 over the separation region, and extends downstream. The result of $h = 38\ mm$ illustrated in Figure 48, Figure 49, and Figure 50 is quite similar, with the SST-SR model overperforming the SST model, fitting better with the LES data. $\beta_{SR}$ is also increased over the separated layer as plotted in Figure 51, but the distribution is more concentrated.

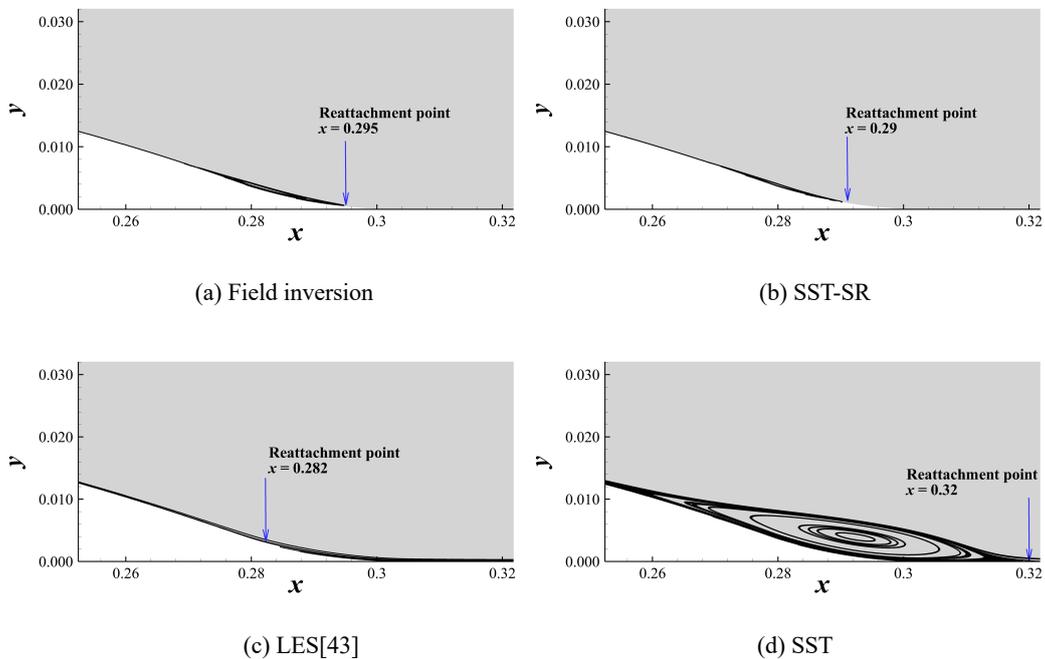

Figure 44. Recirculation zone given by different methods, $h = 26\ mm$

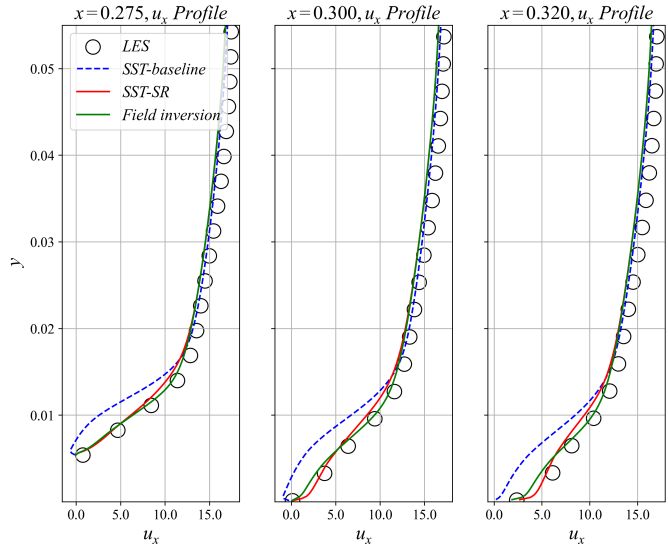

Figure 45. Velocity profiles in the separation region, $h = 26\ mm$.

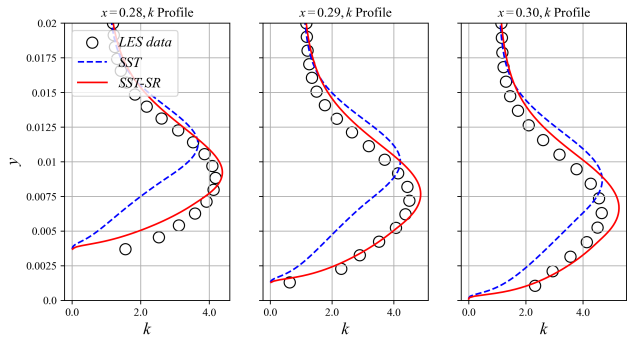

Figure 46. TKE profiles near the separated region, $h = 26\ mm$

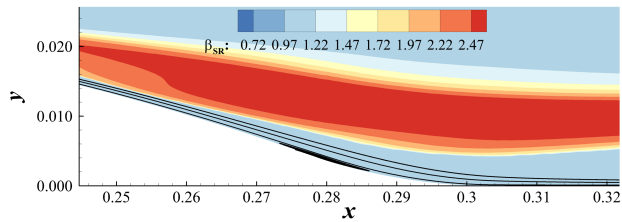

Figure 47. $\beta_{SR}$ contour, $h = 26\ mm$, predicted by the SST-SR model

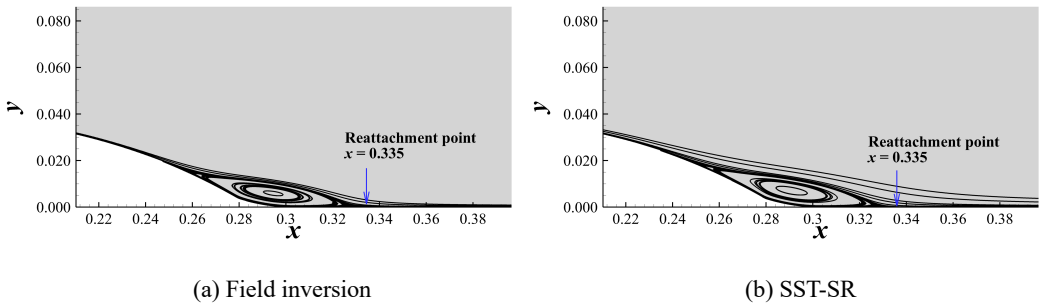

(a) Field inversion  (b) SST-SR

(c) LES[43]

(d) SST

Figure 48. Recirculation zone given by different methods, $h = 38\ mm$.

Figure 49. The velocity profile in the separation region, $h = 38\ mm$.

Figure 50. TKE profiles near the separated region, $h = 38\ mm$

Figure 51. $\beta_{SR}$ contour, $h = 38\ mm$, predicted by the SST-SR model

# Acknowledgment

This work is supported by the Science and Technology Development Program of Jilin Province of China (grant no. 20220301013GX). It is also supported by the National Natural Science Foundation of China (grant nos. 92152301 and 91952302), and the Aeronautical Science Foundation of China (grant no. 2020Z006058002). We are also grateful to Dr. Ping He from Iowa State University for his valuable and effective help in the use of DAFoam and its secondary development.

# Reference

[1]. Rumsey, C. L. "Exploring a Method for Improving Turbulent Separated-Flow Predictions with k-ω Models." *NASA Technical report,* NASA TM-2009-215952, December 1, 2009.

[2]. McConkey, R., Yee, E., and Lien, F.-S. "A Curated Dataset for Data-Driven Turbulence Modelling." *Scientific Data*, Vol. 8, No. 1, 2021, p. 255
doi:https://doi.org/10.1038/s41597-021-01034-2

[3]. Avdis, A., Lardeau, S., and Leschziner, M. "Large Eddy Simulation of Separated Flow over a Two-Dimensional Hump with and without Control by Means of a Synthetic Slot-Jet." *Flow, Turbulence and Combustion*, Vol. 83, No. 3, 2009, pp. 343–370.
doi: https://doi.org/10.1007/s10494-009-9218-y

[4]. Li, H., Zhang, Y., and Chen, H. "Aerodynamic Prediction of Iced Airfoils Based on Modified Three-Equation Turbulence Model." *AIAA Journal*, Vol. 58, No. 9, 2020, pp. 3863–3876.
doi: https://doi.org/10.2514/1.J059206

[5]. Kuntz, M., and Menter, F. Ahmed Car. In *FLOMANIA — A European Initiative on Flow Physics Modelling, Results of the European-Union funded project, 2002 – 2004*, 2006, pp. 335–346.
doi: https://doi.org/10.1007/978-3-540-39507-2

[6]. Li, H., Zhang, Y., and Chen, H. "Optimization of Supercritical Airfoil Considering the Ice-Accretion Effects." *AIAA Journal*, Vol. 57, No. 11, 2019, pp. 4650–4669.
doi: https://doi.org/10.2514/1.J057958

[7]. Li, H., Zhang, Y., and Chen, H. "Numerical Simulation of Iced Wing Using Separating Shear Layer Fixed Turbulence Models." *AIAA Journal*, Vol. 59, No. 9, 2021, pp. 3667–3681.

doi: https://doi.org/10.2514/1.J060143

[8]. Xiao, H., Wang, J.-X., and Ghanem, R. G. "A Random Matrix Approach for Quantifying Model-Form Uncertainties in Turbulence Modeling." *Computer Methods in Applied Mechanics and Engineering*, Vol. 313, 2017, pp. 941–965.

doi: https://doi.org/10.1016/j.cma.2016.10.025

[9]. Xiao, H., Wu, J.-L., Wang, J.-X., Sun, R., and Roy, C. J. "Quantifying and Reducing Model-Form Uncertainties in Reynolds-Averaged Navier-Stokes Simulations: A Data-Driven, Physics-Based Bayesian Approach." *Journal of Computational Physics*, Vol. 324, 2016, pp. 115–136.

doi: https://doi.org/10.1016/j.jcp.2016.07.038

[10]. Karthikeyan Duraisamy, Ze J. Zhang and Anand Pratap Singh. "New Approaches in Turbulence and Transition Modeling Using Data-driven Techniques," AIAA 2015-1284. 53rd AIAA Aerospace Sciences Meeting. January 2015.

doi: https://doi.org/10.2514/6.2015-1284

[11]. Parish, E. J., and Duraisamy, K. "A Paradigm for Data-Driven Predictive Modeling Using Field Inversion and Machine Learning." *Journal of Computational Physics*, Vol. 305, 2016, pp. 758–774.

doi: https://doi.org/10.1016/j.jcp.2015.11.012

[12]. Singh, A. P., Medida, S., and Duraisamy, K. "Machine-Learning-Augmented Predictive Modeling of Turbulent Separated Flows over Airfoils." *AIAA Journal*, Vol. 55, No. 7, 2017, pp. 2215–2227.

doi: https://doi.org/10.2514/1.J055595

[13]. Singh, A. P., and Duraisamy, K. "Using Field Inversion to Quantify Functional Errors in Turbulence Closures." *Physics of Fluids*, Vol. 28, No. 4, 2016, p. 045110.

doi: https://doi.org/10.1063/1.4947045

[14]. Yan, C., Li, H., Zhang, Y., and Chen, H. "Data-Driven Turbulence Modeling in Separated Flows Considering Physical Mechanism Analysis." *International Journal of Heat and Fluid Flow*, Vol. 96, 2022, p. 109004.

doi: https://doi.org/10.1016/j.ijheatfluidflow.2022.109004

[15]. Yan, C., Zhang, Y., and Chen, H. "Data Augmented Turbulence Modeling for Three-Dimensional Separation Flows." *Physics of Fluids*, Vol. 34, No. 7, 2022, p. 075101.

doi: https://doi.org/10.1063/5.0097438

[16]. Yin, Y., Yang, P., Zhang, Y., Chen, H., and Fu, S. "Feature Selection and Processing of Turbulence Modeling Based on an Artificial Neural Network." *Physics of Fluids*, Vol. 32, No. 10, 2020, p. 105117.

doi: https://doi.org/10.1063/5.0022561

[17]. Yin, Y., Shen, Z., Zhang, Y., Chen, H., and Fu, S. "An Iterative Data-Driven Turbulence Modeling Framework Based on Reynolds Stress Representation." *Theoretical and Applied Mechanics Letters*, Vol. 12, No. 5, 2022, p. 100381.

doi: https://doi.org/10.1016/j.taml.2022.100381

[18]. Ho, J., Pepper, N., and Dodwell, T. Probabilistic Machine Learning to Improve Generalisation of Data-Driven Turbulence Modelling. arXiv preprint arXiv:2301.09443, 2023.

doi: https://doi.org/10.48550/arXiv.2301.09443

[19]. Christopher L. Rumsey, Gary N. Coleman and Li Wang. "In Search of Data-Driven Improvements to RANS Models Applied to Separated Flows," AIAA 2022-0937. AIAA SCITECH 2022 Forum. January 2022.

doi: https://doi.org/10.2514/6.2022-0937

[20]. Spalart, P. "An Old-Fashioned Framework for Machine Learning in Turbulence Modeling." NASA 2022 Symposium on Turbulence Modeling: Roadblocks, and the Potential for Machine Learning. July 2022.

[21]. Miles Cranmer. (2020). MilesCranmer/PySR v0.2 (v0.2). Zenodo. September 2020

doi: https://doi.org/10.5281/zenodo.4041459

[22]. Cranmer, Miles, et al. "Discovering symbolic models from deep learning with inductive biases." Advances in Neural Information Processing Systems 33 (2020): 17429-17442. ISBN: 9781713829546.

[23]. Cranmer, Miles. "Interpretable machine learning for science with PySR and SymbolicRegression. jl." arXiv preprint arXiv:2305.01582, 2023.

doi: https://doi.org/10.48550/arXiv.2305.01582

[24]. Sahoo, Subham, Christoph Lampert, and Georg Martius. "Learning equations for extrapolation and control." International Conference on Machine Learning. PMLR, 2018.

[25]. Schmelzer, M., Dwight, R. P., and Cinnella, P. "Discovery of Algebraic Reynolds-Stress Models Using Sparse Symbolic Regression." *Flow, Turbulence and Combustion*, Vol. 104, Nos. 2–3, 2020, pp. 579–603.

doi: https://doi.org/10.1007/s10494-019-00089-x.

[26]. Xie, H., Zhao, Y., and Zhang, Y. "Data-Driven Nonlinear K-L Turbulent Mixing Model via Gene Expression Programming Method." *Acta Mechanica Sinica*, Vol. 39, No. 2, 2023, p. 322315.

doi: https://doi.org/10.1007/s10409-022-22315-x.

[27]. Menter, F. R., Kuntz, M., and Langtry, R. "Ten Years of Industrial Experience with the SST Turbulence Model." *Heat and Mass Transfer*, Vol. 4, No. 1, 2003, pp. 625–632.

[28]. Antti Hellsten. "Some improvements in Menter's k-omega SST turbulence model," AIAA 1998-2554. 29th AIAA, Fluid Dynamics Conference. June 1998.

doi: https://doi.org/10.2514/6.1998-2554

[29]. Gill, P. E., Murray, W., and Saunders, M. A. "SNOPT: An SQP Algorithm for Large-Scale Constrained Optimization." *SIAM Review*, Vol. 47, No. 1, 2005, pp. 99–131.

doi: https://doi.org/10.1137/S0036144504446096

[30]. Jameson, A., and Kim, S., "Reduction of the Adjoint Gradient Formula for Aerodynamic Shape Optimization Problems," *AIAA Journal*, Vol. 41, No. 11, 2003, pp. 2114–2129.

doi: https://doi.org/10.2514/2.6830

[31]. Duffy, Austen C. "An introduction to gradient computation by the discrete adjoint method." *Technical report*, Florida State University, 2009.

[32]. Griewank, A., & Walther, A. Evaluating Derivatives. Society for Industrial and Applied Mathematics. 2008.

doi: https://doi.org/10.1137/1.9780898717761

[33]. He, P., Mader, C. A., Martins, J. R. R. A., and Maki, K. J. "DAFoam: An Open-Source Adjoint Framework for Multidisciplinary Design Optimization with OpenFOAM." *AIAA Journal*, Vol. 58, No. 3, 2020, pp. 1304–1319.

doi: https://doi.org/10.2514/1.J058853.

[34]. Ping He, Charles A. Mader, Joaquim R. R. A. Martins and Kevin Maki. "An Object-oriented Framework for Rapid Discrete Adjoint Development using OpenFOAM," AIAA 2019-1210. AIAA Scitech 2019 Forum. January 2019.

doi: https://doi.org/10.2514/6.2019-1210

[35]. He, P., Mader, C. A., Martins, J. R. R. A., and Maki, K. J. "An Aerodynamic Design Optimization Framework Using a Discrete Adjoint Approach with OpenFOAM." *Computers & Fluids*, Vol. 168, 2018, pp. 285–303.

doi: https://doi.org/10.1016/j.compfluid.2018.04.012.

[36]. Omid Bidar, Ping He, Sean Anderson and Ning Qin. "An Open-Source Adjoint-based Field Inversion Tool for Data-driven RANS modelling," AIAA 2022-4125. AIAA AVIATION 2022 Forum. June 2022.

doi: https://doi.org/10.2514/6.2022-4125

[37]. Omid Bidar, Ping He, Sean Anderson and Ning Qin. "Turbulent Mean Flow Reconstruction Based on Sparse Multi-sensor Data and Adjoint-based Field Inversion," AIAA 2022-3900. AIAA AVIATION 2022 Forum. June 2022.

doi: https://doi.org/10.2514/6.2022-3900

[38]. Jasak, H., Jemcov, A., and Tukovic, Z. OpenFOAM: A C++ Library for Complex Physics Simulations. In *International workshop on coupled methods in numerical dynamics*, 2007, pp. 1–20.

[39]. Sagebaum, M., Albring, T., and Gauger, N. R. "High-Performance Derivative Computations Using CoDiPack." *ACM Transactions on Mathematical Software*, Vol. 45, No. 4, 2019, p. 38:1-38:26.

doi: https://doi.org/10.1145/3356900.

[40]. Pope, S. B. "A More General Effective-Viscosity Hypothesis." *Journal of Fluid Mechanics*, Vol. 72, No. 02, 1975, p. 331.

doi: https://doi.org/10.1017/S0022112075003382.

[41]. Bentaleb, Y., Lardeau, S., and Leschziner, M. A. "Large-Eddy Simulation of Turbulent Boundary Layer Separation from a Rounded Step." *Journal of Turbulence*, Vol. 13, 2012, p. N4.

doi: https://doi.org/10.1080/14685248.2011.637923.

[42]. Speziale, C. G., Abid, R., and Anderson, E. C., "Critical Evaluation of Two-Equation Models for Near-Wall Turbulence," *AIAA Journal*, Vol. 30, No. 2, 1993, pp. 324–331.

doi: https://doi.org/10.2514/3.10922

[43]. Matai, R., and Durbin, P. "Large-Eddy Simulation of Turbulent Flow over a Parametric Set of Bumps." *Journal of Fluid Mechanics*, Vol. 866, 2019, pp. 503–525.

doi: https://doi.org/10.1017/jfm.2019.80.

[44]. Xiao, H., Wu, J.-L., Laizet, S., and Duan, L. "Flows over Periodic Hills of Parameterized Geometries: A Dataset for Data-Driven Turbulence Modeling from Direct Simulations." *Computers & Fluids*, Vol. 200, 2020, p. 104431.

doi: https://doi.org/10.1016/j.compfluid.2020.104431.

[45]. Lienhart, H., C. Stoots, and S. Becker. "Flow and turbulence structures in the wake of a simplified car model (ahmed modell)." New Results in Numerical and Experimental Fluid Mechanics III: Contributions to the 12th STAB/DGLR Symposium Stuttgart, Germany 2000. Springer Berlin Heidelberg, 2002.

doi: https://doi.org/10.1007/978-3-540-45466-3_39

[46]. Haase, W., Ed. *FLOMANIA: A European Initiative on Flow Physics Modelling: Results of the European-Union Funded Project, 2002-2004*. Springer, Berlin ; New York, 2006.

doi: https://doi.org/10.1007/978-3-540-39507-2

[47]. Guilmineau, E., Deng, G. B., Leroyer, A., Queutey, P., Visonneau, M., and Wackers, J. Aessessment of RANS and DES Methods for The Ahmed Body. Presented at the VII European Congress on Computational Methods in Applied Sciences and Engineering, Crete Island, Greece, 2016.

doi: http://dx.doi.org/10.7712/100016.1860.10023

[48]. Wieghardt, K., and Tillmann, W., "On the Turbulent Friction Layer for Rising Pressure," *NASA Technical report*, NACA-TM-1314. October 1, 1951